\documentclass[twocolumn,reprint,amsfonts,amsmath,amssymb,nofootinbib,byrevtex]{revtex4-1}

\usepackage{graphicx}
\usepackage{dcolumn}
\usepackage{bm}
\usepackage{hyperref}
\usepackage[caption=false]{subfig}
\usepackage{multirow}

\usepackage{rotating}
\usepackage{tabularx}

\usepackage{eurosym}
\usepackage{color}
\usepackage{soul}

\newenvironment{Tabular}[2][1]
  {\def\arraystretch{#1}\tabular{#2}}
  {\endtabular}

\begin{document}
\title{Measurement of atmospheric neutrino oscillations with very large volume neutrino telescopes}
\author{J.~P. Y\'{a}\~{n}ez}
\thanks{\href{mailto:juan.pablo.yanez@desy.de}{juan.pablo.yanez@desy.de}}
\affiliation{\footnotesize DESY, D-15735 Zeuthen, Germany}
\author{A. Kouchner}
\thanks{\href{mailto:kouchner@apc.univ-paris7.fr}{kouchner@apc.univ-paris7.fr}}
\affiliation{\footnotesize Laboratoire AstroParticule et Cosmologie (APC), Universit\'e Paris Diderot, CNRS/IN2P3, CEA/IRFU, Observatoire de Paris, Sorbonne Paris Cit\'e, 75205 Paris, France} 

\date{\today}

\begin{abstract}
\textsc{Abstract$-$}Neutrino oscillations have been probed during the last few decades using multiple neutrino sources and experimental set-ups. In the recent years, very large volume neutrino telescopes have started contributing to the field. First ANTARES and then IceCube have relied on large and sparsely instrumented volumes to observe atmospheric neutrinos for combinations of baselines and energies inaccessible to other experiments. Using this advantage, the latest result from IceCube starts approaching the precision of other established technologies, and is paving the way for future detectors, such as ORCA and PINGU. These new projects seek to provide better measurements of neutrino oscillation parameters, and eventually determine the neutrino mass ordering. The results from running experiments and the potential from proposed projects are discussed in this review, emphasizing the experimental challenges involved in the measurements.
\end{abstract}
\maketitle

\section{Introduction}

Massive, mixed neutrinos, inferred from the phenomenon of oscillations, remain until this day the only physics found beyond the original formulation of the Standard Model. While the Standard Model can be extended to account for these experimental facts, precise measurements of the parameters involved in the phenomenon are necessary to constrain the different theories that attempt to explain it.

The current knowledge favors the existence of three active neutrinos (flavor eigenstates $\nu_e,\,\nu_\mu,\,\nu_\tau$) whose mixing can be fully determined by the PMNS\footnote{Pontecorvo-Maki-Nakagawa-Sakata} matrix $U$. The matrix is often parameterized as the product of three rotation matrices, related to the mixing angles $\theta_{12}$, $\theta_{13}$ and $\theta_{23}$, and a complex CP phase $\delta$:
\begin{align} \label{eq:full_matrix}
U =
 \begin{pmatrix}
  1 & 0 & 0 \\
  0 & c_{23} & s_{23} \\
  0 & -s_{23} & c_{23}
 \end{pmatrix} \times
 &\begin{pmatrix}
  c_{13} & 0 & e^{-i\delta}s_{13} \\
  0 & 1 & 0 \\
  -e^{i\delta}s_{13} & 0 & c_{13}
 \end{pmatrix}\times\\
 &\begin{pmatrix}
  c_{12} & s_{12} & 0 \\
  -s_{12} & c_{12} & 0 \\
  0 & 0 & 1
 \end{pmatrix}\times
 \begin{pmatrix}
  e^{i\rho_1} & 0 & 0 \\
  0 & e^{i\rho_2} & 0 \\
  0 & 0 & 1
 \end{pmatrix},\nonumber
\end{align}
where $c_{ij} \equiv \cos\theta_{ij}$ and $s_{ij} \equiv \sin\theta_{ij}$. The last matrix in the multiplication does not affect neutrino oscillations, and only exists if neutrinos are Majorana particles \cite{pdg2014}.

The three angles that determine the mixing matrix are known to a precision of 10\% or better \cite{Forero:2014bxa,Gonzalez-Garcia:2014bfa,T2K_latest}. The absolute differences of the square of the masses (mass splittings $\Delta m^2_{ij} = m_i^2 - m_j^2$  with $i,j=1,2,3$), which play a role in oscillations, are known to a precision better than 5\%.  While the sign of the mass splitting between the states 1-2 ($m_2>m_1$) is known from matter effects in solar neutrino oscillations, the relative difference between $m_3$ and $m_1$ remains unknown. The relative values of the neutrino masses are commonly referred to as the neutrino mass ordering (NMO)\footnote{The determination of the mass ordering is sometimes confused with the determination of the neutrino ``mass hierarchy'', which requires additional information on the absolute scale of the neutrino masses \cite{winter_nu14}.}, which has two possible options: the normal ordering (NO), with $m_1 < m_2 < m_3$, and the inverted ordering (IO), with $m_3 < m_1 < m_2$. The determination of the NMO is important as the parameter can discriminate between flavor symmetry models \cite{NMH_implications}. Also, the sensitivity of experiments attempting to determine the neutrino nature depend on the NMO \cite{neutrinoless1, neutrinoless2}. Finally, knowing the NMO would help to measure the value of the $\delta$ phase, which in turn would be an important step forward towards solving the fundamental question of the prevalence of matter over anti-matter in the Universe. Better measurements of all the parameters involved in neutrino oscillations are therefore necessary to understand if the current model is correct and how to incorporate it to the Standard Model.

In this view, atmospheric neutrinos remain a promising tool for studying oscillations: they cover a wide energy range, from MeV to TeV, and can reach a detector after traveling distances from a few to about 12700\,km when they cross the Earth. No man-made beam covers a similar parameter space. However, the flux strongly decreases with energy. Detecting it implies building large detectors, such as very large volume neutrino telescopes (VLVNTs) to study atmospheric neutrino oscillations.

In the recent years first ANTARES and then IceCube/DeepCore have proven that these studies are feasible by analyzing interactions of neutrinos with energy as low as 15\,GeV. The current result from IceCube on $\sin^2\theta_{23}$ and $\Delta m_{32}^2$ reaches a precision which is only a factor of three to four times less stringent than global fits which combine all available data \cite{Forero:2014bxa,Gonzalez-Garcia:2014bfa,icecube_yanez}. Building upon the success of these studies, proposals of new, more densely instrumented telescopes, as extensions or part of new projects, have appeared. PINGU and ORCA aim to improve the precision of these measurements and reduce the energy threshold to a few GeV, where matter effects are strong, and use these effects to measure the NMO.

This review begins by covering, in Section~\ref{sec:atm_osc}, the current knowledge on atmospheric neutrinos, and how oscillations affect their flux. Section~\ref{sec:vlvnt} describes the design and operation of VLVNTs. Special attention is paid to relevant sources of uncertainty. The neutrino oscillation results produced by these experiments until this date are covered in Section~\ref{sec:results}. Section~\ref{sec:future} discusses the possible studies with future detectors, and a short summary is given in Section~\ref{sec:summary}.
\section{Atmospheric neutrino oscillations}
\label{sec:atm_osc}

The flux of atmospheric neutrinos in the energies relevant for a VLVNT, together with how the flux is modified by neutrino oscillations for those neutrinos that cross the Earth, are the topics covered in this Section.

\subsection{A neutrino beam from cosmic rays}
Cosmic rays (CR) continuously arrive at the Earth from all directions and interact with nuclei in the atmosphere at altitudes of about 25\,km above sea level and initiate showers of particles. During the shower development, charged mesons are produced that eventually decay in comparable numbers of muons and neutrinos:
\begin{alignat}{2}
  \label{eq:nu_production}
  CR+N \rightarrow X + &\pi^\mp,\,K^\mp\,, &\nonumber\\
  &\pi^-,\,K^- \rightarrow \;&\mu^- + \bar{\nu}_\mu\,, \nonumber\\
  &\pi^+,\,K^+ \rightarrow \;&\mu^+ + \nu_\mu\,, \\
  & & \mu^- \rightarrow \;e^-& + \bar{\nu}_e + \nu_\mu\,, \nonumber \\
  & & \mu^+ \rightarrow \;e^+& + \nu_e + \bar{\nu}_\mu\,. \nonumber
\end{alignat}

The atmospheric muons produced in air showers can travel long distances before they decay. They are able to penetrate deep into the Earth, depending on their energy and the material that they are crossing \cite{pdg2014}, and constitute the dominant background in the measurement of atmospheric neutrinos.

\begin{figure}[bth!]
  \centering
  \includegraphics[ width=0.48\textwidth]{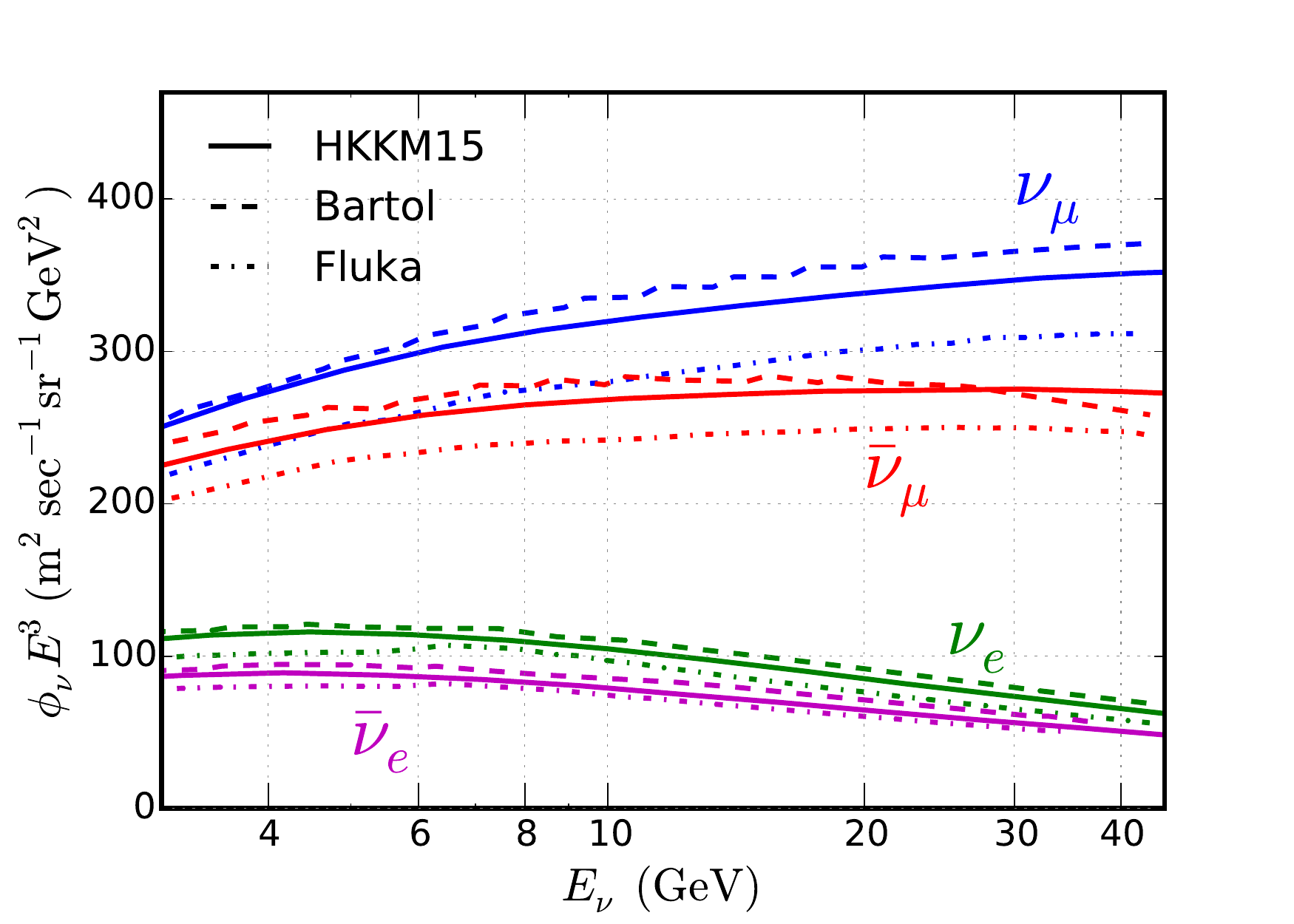}
  \caption{Comparison of predicted atmospheric neutrino fluxes per flavor for the energy range relevant for neutrino oscillation measurements with VLVNT. Reproduced from \cite{honda2011,honda2015}.}
  \label{fig:atmnu_flux}
\end{figure}

\begin{figure}[bth!]
  \centering
  \includegraphics[width=0.48\textwidth]{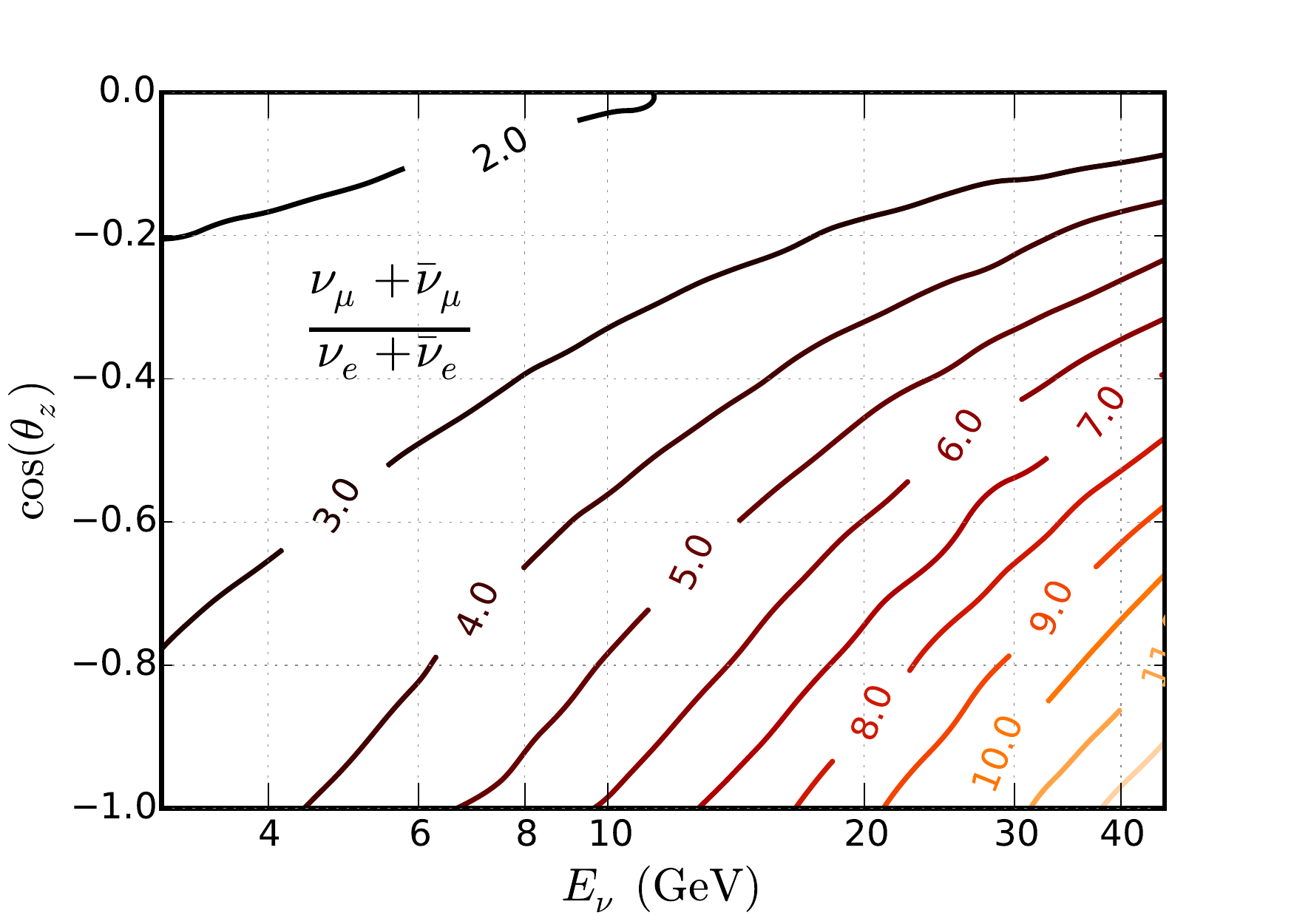}
  \caption{Iso-contours of the ratio of $\left(\nu_\mu + \bar{\nu}_\mu \right) / \left(\nu_e + \bar{\nu}_e \right)$ as a function of energy and neutrino arrival direction for neutrinos that cross the Earth, as predicted by the latest HKKM model \cite{honda2015}. The ratio is nearly up-down symmetric.}
  \label{fig:atmnu_ratios}
\end{figure}

Atmospheric neutrinos have been measured over a wide energy range \cite{nusex_flux, frejus_spectrum, soudan_flux, imb_flux, kamiokande_flux, sk_flux, amanda_spectrum, antares_spectrum, icecube_unfolding} and multiple models predict their flux \cite{honda2011, Barr:2004br,Battistoni:2002ew} (see Fig.~\ref{fig:atmnu_flux}). The most noticeable difference between the models is the absolute flux, which changes by up to 20\% both for electron and muon (anti)neutrinos. Apart from that, the models agree that atmospheric neutrinos follow a power-law energy spectrum with a spectral index close to 3 in the energy range $E_\nu= [3-100]$\,GeV. Measurements from \cite{frejus_spectrum} estimate an uncertainty of $\pm0.04$ on the spectral index of atmospheric neutrinos. A similar uncertainty has been also derived from varying the underlying cosmic ray model in neutrino flux calculations \cite{Barr:2006it}.

Muon neutrinos dominate the flux, and also have the hardest spectral index (see Fig.~\ref{fig:atmnu_flux}). Since electron (anti)neutrinos mainly come from muons that lose energy before decaying (see Eq.\,\ref{eq:nu_production}) their spectral index is softer, and their relative contribution to the total neutrino flux depends on energy and direction. The direction-averaged flux of $\bar{\nu}_\mu$ is between 1.1 and 1.3 times smaller than that of $\nu_\mu$, depending on the energy. The electron (anti)neutrino direction-averaged flux at a few GeV is about 2.5 times smaller than its muon (anti)neutrino counterpart. Figure~\ref{fig:atmnu_ratios} shows iso-contours of the neutrino flux flavor ratio as a function of energy and zenith angle for neutrinos that cross the Earth. The flux difference between $\nu_\mu$ and $\nu_e$ grows both with energy and $|\cos\theta_z|$. Already at 40\,GeV the direction-averaged ratio is close to four.

Above 10\,GeV the neutrino flux as a function of zenith angle is almost symmetric around $\cos \theta_z = 0$. The angular dependence of the flux at the detection site is influenced by hadronization processes, the local atmospheric density and geomagnetic effects at the interaction point. The uncertainties associated to these processes result in energy-dependent modeling errors of the arrival zenith angle by up to 20\,\% on the ratio $\nu/\bar\nu$ for muon neutrinos and 8\,\% for electron neutrinos \cite{Barr:2006it}.

\subsection{Neutrino oscillations at $E_\nu \geq 5$\,GeV}

The flux of atmospheric neutrinos at a detection site is modified by oscillations. The oscillations that act over the $L/E$ parameter space accessible with atmospheric neutrinos that cross the Earth ($L/E \sim 10^1 - 10^3$\,km/GeV) are mainly driven by the large mass splitting, $\Delta m^2_{32} \simeq \Delta m^2_{31}$ and the mixing angles $\theta_{13}$, $\theta_{23}$. These are therefore parameters that VLVNTs are sensitive to.

Neutrinos propagating in matter are subject to a potential due to coherent forward scattering with the particles in the medium \cite{Wolfenstein:1977ue}. For explanatory purposes, we consider the case of neutrinos traveling through matter with constant electron density that results in a potential $A = \pm 2\sqrt{2}\,G_F\, n_e(x)\,E_\nu$, where $G_F$ is the Fermi constant, and the plus (minus) sign corresponds to neutrinos (antineutrinos). Computation of neutrino oscillation probabilities for the relevant energies has been done in \cite{Choubey:2005zy}, from where we take the approximations for the $\nu_\mu$ to $\nu_e$ transition, given by 
\begin{equation}
P_{\mu e} \simeq \sin^2\theta_{23} \sin^2 2\theta^M_{13} \sin^2\left[ \varDelta^M \frac{L}{4E} \right],
\label{eq:pmue}
\end{equation}
while the survival probability of $\nu_\mu$ is a somewhat more complicated expression,
\begin{align}
\label{eq:pmumu}
P_{\mu \mu} \simeq 1&-\sin^2\theta_{13}^M \sin^2 2\theta_{23}  \sin^2\left[ \left(\varDelta - \varDelta^M + A \right)\frac{L}{8E}\right]\\ \nonumber
&-\cos^2\theta_{13}^M\sin^22\theta_{23} \sin^2\left[\left(\varDelta + \varDelta^M + A \right)\frac{L}{8E}\right]\\ \nonumber
&- \sin^4 \theta_{23} \sin^2 2\theta_{13}^M \sin^2\left[ \varDelta^M  \frac{L}{4E}\right],
\end{align}
and the transitions to $\nu_\tau$ are simply 
\begin{equation}
P_{\mu \tau} \simeq 1 - P_{\mu e} - P_{\mu \mu}.
\label{eq:pmutau}
\end{equation}
In these expressions $\varDelta \equiv \Delta m^2_{31}$ and $\varDelta^M$ is the effective mass splitting in matter, given by
\begin{equation}
\varDelta^M \simeq \sqrt{\left(\Delta m_{31}^2 \cos 2\theta_{13} - A \right)^2 + \left(\Delta m_{31}^2 \sin 2\theta_{13}\right)^2}.
\label{eq:mass_matter}
\end{equation}
The superscript $M$ also accompanies $\theta_{13}$, whose effective value in matter is
\begin{equation}
\label{eq:theta_matter}
\sin2\theta^M_{13} \simeq \dfrac{\Delta m^2_{31} \, \sin 2 \theta_{13}}{\varDelta^M}.
\end{equation}

The mixing angle $\theta_{23}$ is known to be close to maximal ($\sim\pi/4$), and $|\Delta m^2_{31}|$ is of the order of $10^{-3}$\,eV$^2$ \cite{pdg2014}. The angle $\theta_{13}$ has been recently measured and found to be small but non-zero \cite{dayabay_results,reno_results,dchooz_results}. It is then the case that $\theta^M_{13}$ can acquire any value, depending on the neutrino energy and the electron density of the material being crossed, as shown in Eq.~\ref{eq:theta_matter}. For a low electron density or neutrino energy, the parameters (and equations) in vacuum are recovered. A particularly interesting case appears when $A =\Delta m^2_{31}\cos2\theta_{13}$, which gives $\theta^M_{13} = \pi/4$, maximizing the mixing between states 1-3, i.e. a resonance appears \cite{smirnov_resonance}. The effective mass splitting acquires its minimum value under this condition, and is reduced by a factor $\sin 2\theta_{13}$.

The resonance that leads to maximal 1-3 mixing can only happen if the potential $A$ and the mass difference $\Delta m^2_{31}$ have the same sign, so for neutrinos in the case of NO and anti-neutrinos in the case of IO. Identifying whether the resonance takes place in neutrinos or antineutrinos is a way to identify the NMO.

For $A \gg \Delta m^2_{31}\cos2\theta_{13}$ a saturation effect occurs, where the effective angle in matter goes to $\pi/2$ and the effective mass splitting is then well approximated by $A$. In the saturated regime transitions of the type $\nu_e \rightarrow \nu_\mu$, given in Eq.~\ref{eq:pmue}, are suppressed by the factor $\sin^22\theta_{13}^M$. The effective matter parameters also modify $P_{\mu\mu}$ by making the last two terms in Eq.~\ref{eq:pmumu} go to zero, resulting in the simpler expression
\begin{equation}
P_{\mu\mu} = 1 - \sin^2 2\theta_{23} \sin^2\left[\varDelta \frac{L}{4E}\right]\,,
\label{eq:osc_musimple}
\end{equation}
with all the oscillated $\nu_\mu$ turning into $\nu_\tau$.

The CP-violating phase $\delta$ is not present in the approximate formulas shown. The reason is that the parameter $\delta$ always appears in oscillation probabilities accompanied by a factor $\Delta m^2_{21}/ \Delta m^2_{31}$, which suppresses its contribution \cite{Freund:2001pn}. Note, however, that the approximations presented here serve the purpose of explaining the main features of neutrino oscillations in matter. Figures contained in this review, as well as the latest data analyses discussed, use numerical calculations of oscillation probabilities that do not rely on simplified analytical expressions.

\subsection{An oscillating atmospheric neutrino flux}
\begin{figure*}[bht!]
  \centering
  \includegraphics[width=0.98\textwidth]{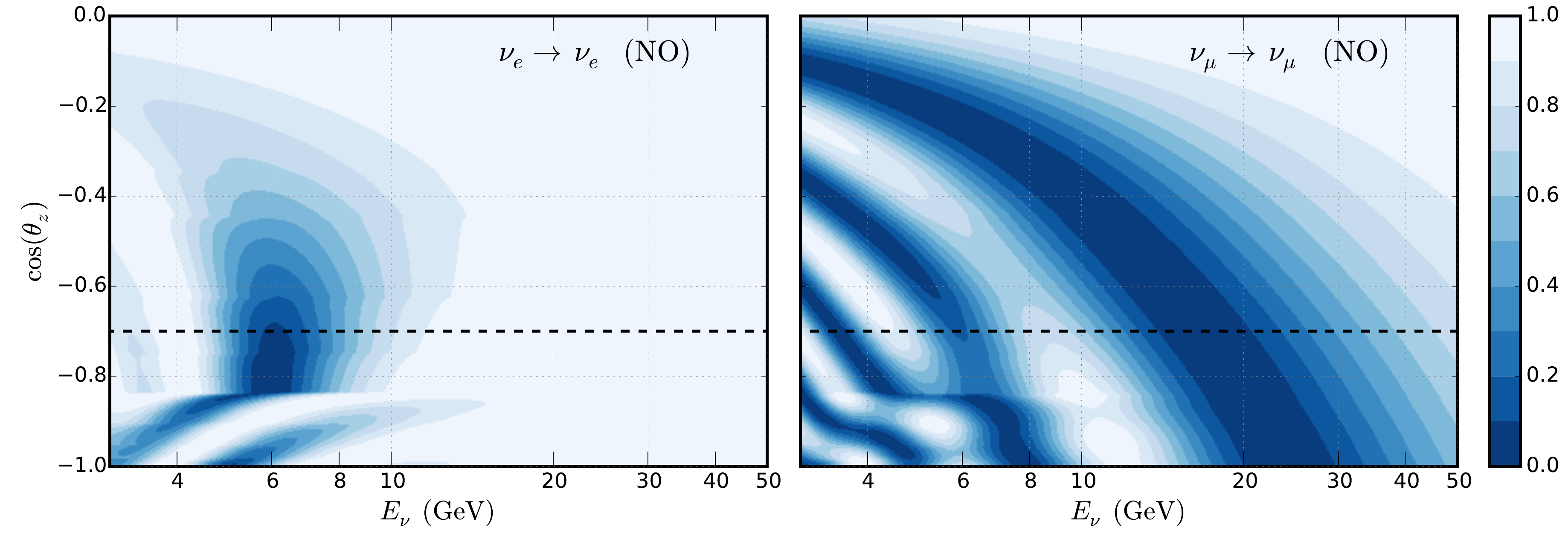}
  \caption{\label{fig:osc_prob1}Survival probabilities for $\nu_e$ (left) and $\nu_\mu$ (right) as a function of neutrino energy and arrival direction for Earth crossing trajectories affected by oscillations ($\cos(\theta_z) \leq 0$). Calculated using the values in \cite{Gonzalez-Garcia:2014bfa} assuming a normal mass ordering. Resonant matter effects produce the large disappearance of $\nu_e$ around 6\,GeV and $\cos\theta_z \sim -0.8$, as well as the discontinuities on the survival pattern of $\nu_\mu$ below 15\,GeV. The abrupt changes observed at $\cos\theta_z\sim-0.85,\,-0.45$ are due to sharp jumps in the electron density profile of the Earth. The dashed line indicates the connection between these figures and Fig.~\ref{fig:osc_prob2}.}
\end{figure*}

The atmospheric neutrinos under consideration, of a few GeV, are mostly $\nu_\mu + \bar{\nu}_\mu$ produced around a height of 25\,km in the atmosphere, where the matter density is low enough to be approximated as vacuum. For most production angles the neutrinos proceed to cross the Earth, which has a non-negligible matter density.

Earth's matter profile can be well explained as concentric shells, each one with a constant density \cite{prem}. To study the transitions that take place, consider the oscillation parameters from \cite{Gonzalez-Garcia:2014bfa} and the electron number density of the mantle, $n_e = 2.5$\,cm$^{-3}N_A$, where $N_A$ is Avogadro's number. Neutrinos crossing the mantle experience the resonance around $E_\nu \simeq 6$\,GeV (see Eq.~\ref{eq:theta_matter}), while the saturation condition $A \gg \Delta m^2_{31}\cos2\theta_{13}$ is fulfilled already at $E_\nu \sim 12$\,GeV. Neutrinos measured by VLVNTs then experience oscillations either in the resonant or saturated regime, depending on the energy threshold of the detector.

Another interesting effect takes place on neutrinos that cross the Earth's core. These neutrinos experience a symmetric electron density profile that changes abruptly. For the right combination of neutrino energy and electron densities, a so-called \textit{parametric resonance} can appear \cite{param1, param2, param3, param4}. The effect, however, is not the dominant one at the energies at which future projects (Section~\ref{sec:future}) will be sensitive to. 

In the saturated regime atmospheric neutrino oscillations are independent of the mass ordering, dominated by $\nu_\mu \rightarrow \nu_\tau$ transitions and well described by Eq.\,\ref{eq:osc_musimple}. Near the resonance condition transitions involving electron (anti)neutrinos also play a role, and patterns become complex. Figure\,\ref{fig:osc_prob1} shows the survival probabilities of $\nu_e$ and $\nu_\mu$ for neutrinos and normal mass ordering. The original electron neutrino flux is expected to fully disappear due to matter effects over $E_\nu = [5, 8]$\,GeV and $\cos\theta_z = [-0.9, -0.5]$. The suppression of these oscillations due to saturation can be observed at about 10\,GeV. The survival probability of $\nu_\mu$ shows abrupt changes that are due to the effects of matter. Muon neutrinos oscillate even if the resonance conditions are not fulfilled, which makes the effects of the resonance less obvious than for electron neutrinos. Resonant matter effects appear in the $\nu_\mu$ survival probability as modifications on the otherwise smooth and periodic disappearance pattern, as shown in Fig.~\ref{fig:osc_prob1}. Saturation is reached above 15\,GeV and the survival probability becomes smooth.

\begin{figure}[bht!]
        \begin{subfloat}
                \centering
                \includegraphics[width=0.45\textwidth]{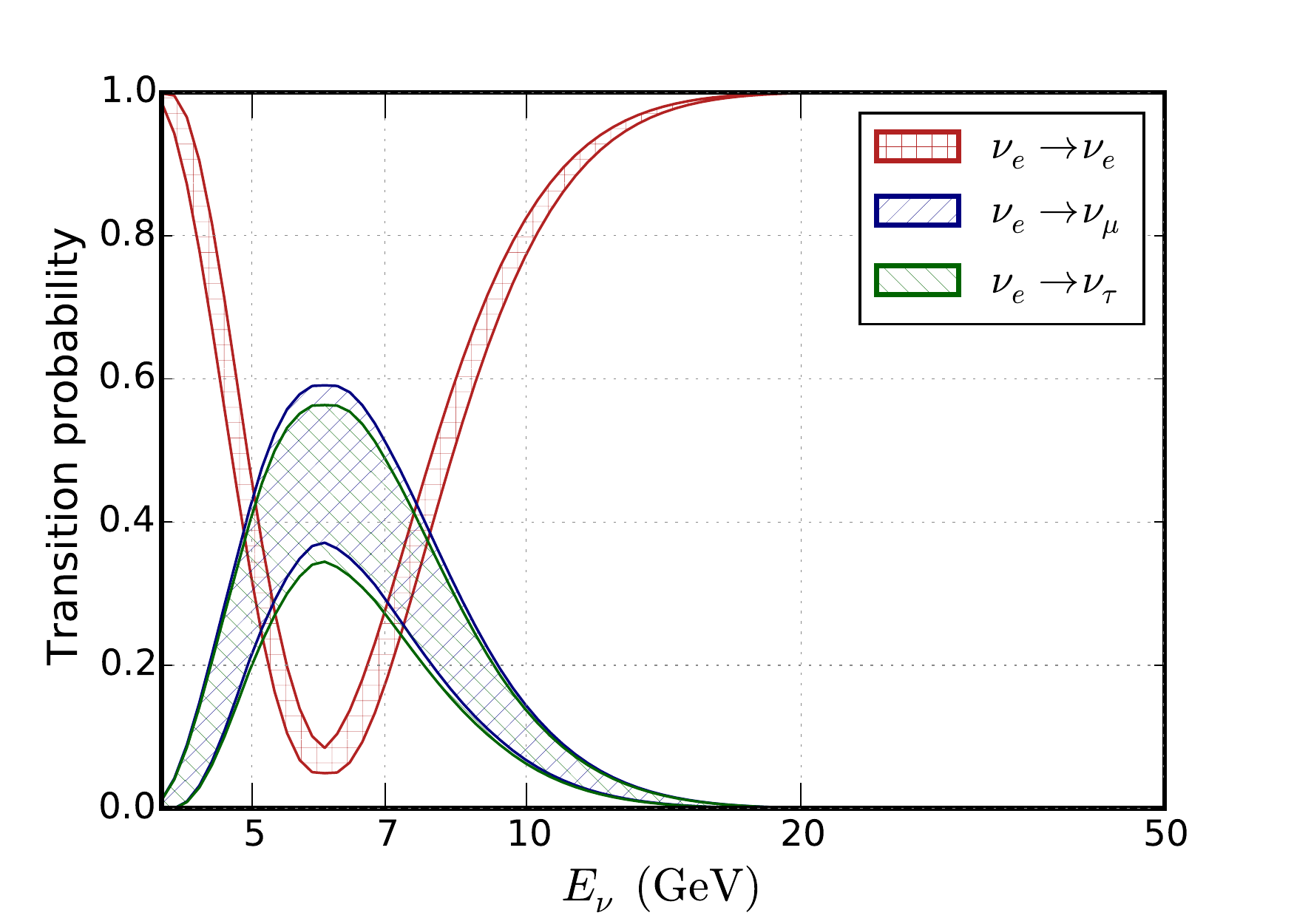}
        \end{subfloat}
        \begin{subfloat}
                \centering
                \includegraphics[width=0.45\textwidth]{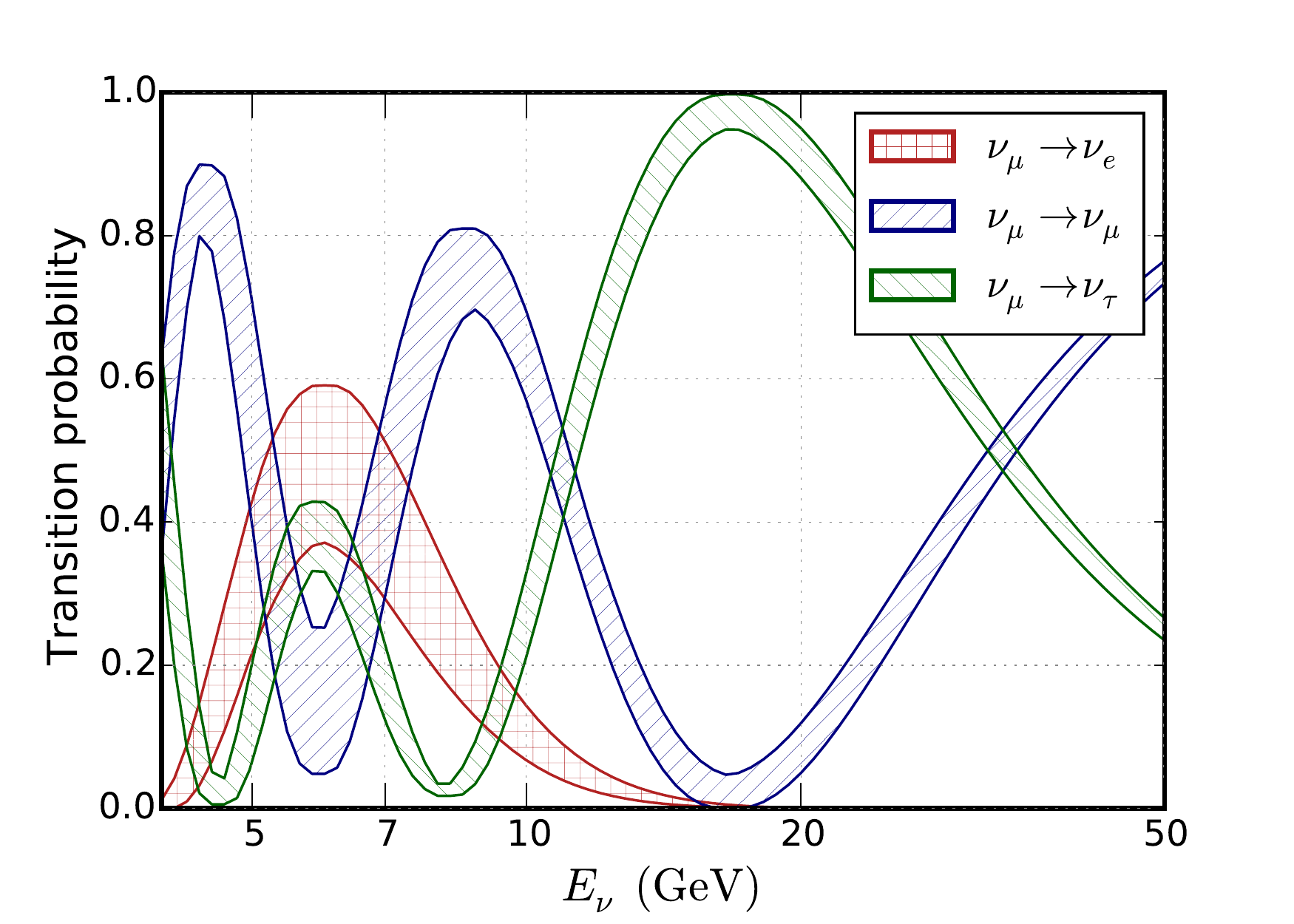}
        \end{subfloat}
        \caption{Transition probabilities for electron (top) and muon (bottom) neutrinos that arrive at a detector from $\cos\theta_z = -0.7$ (mantle-crossing trajectory marked by a dashed line in Fig.~\ref{fig:osc_prob1}). The bands encompass the results of the calculation once the uncertainties on the oscillation parameters from \cite{Gonzalez-Garcia:2014bfa} are included. Normal mass ordering is assumed. If the resonance was absent (inverted mass ordering or transitions for antineutrinos) the top figure would show oscillations with amplitudes smaller than 0.1, while the bottom would show transitions only between muon and tau neutrinos.}
        \label{fig:osc_prob2}
\end{figure}

Figure~\ref{fig:osc_prob2} shows the transition probabilities of $\nu_e$ and $\nu_\mu$ into different flavors for the arrival direction $\cos\theta_z = -0.7$ assuming a normal mass ordering. They correspond to a one-dimensional projection of Fig.~\ref{fig:osc_prob1} along the dashed line. The bands demonstrate how the uncertainties on the oscillation parameters impact the expected probabilities. For $\nu_e$ it is easy to observe the same disappearance as in Fig.~\ref{fig:osc_prob1}, with neutrinos oscillating equally into $\nu_\mu$ and $\nu_\tau$. Transitions of $\nu_\mu$ to other flavors are complicated by matter effects, which open the $\nu_\mu \leftrightarrow \nu_e$ channel and thus modify the survival probability of $\nu_\mu$.

Measurements of neutrino fluxes above the saturation energy of about 15\,GeV are largely independent of $\theta_{13}$, the neutrino/antineutrino admixture of the sample and the ordering of neutrino masses. They provide excellent data for determining $\sin^2\theta_{23}$ as well as $|\Delta m^2_{31}|$.

\begin{figure}[bht!]
  \centering
  \includegraphics[width=0.49\textwidth]{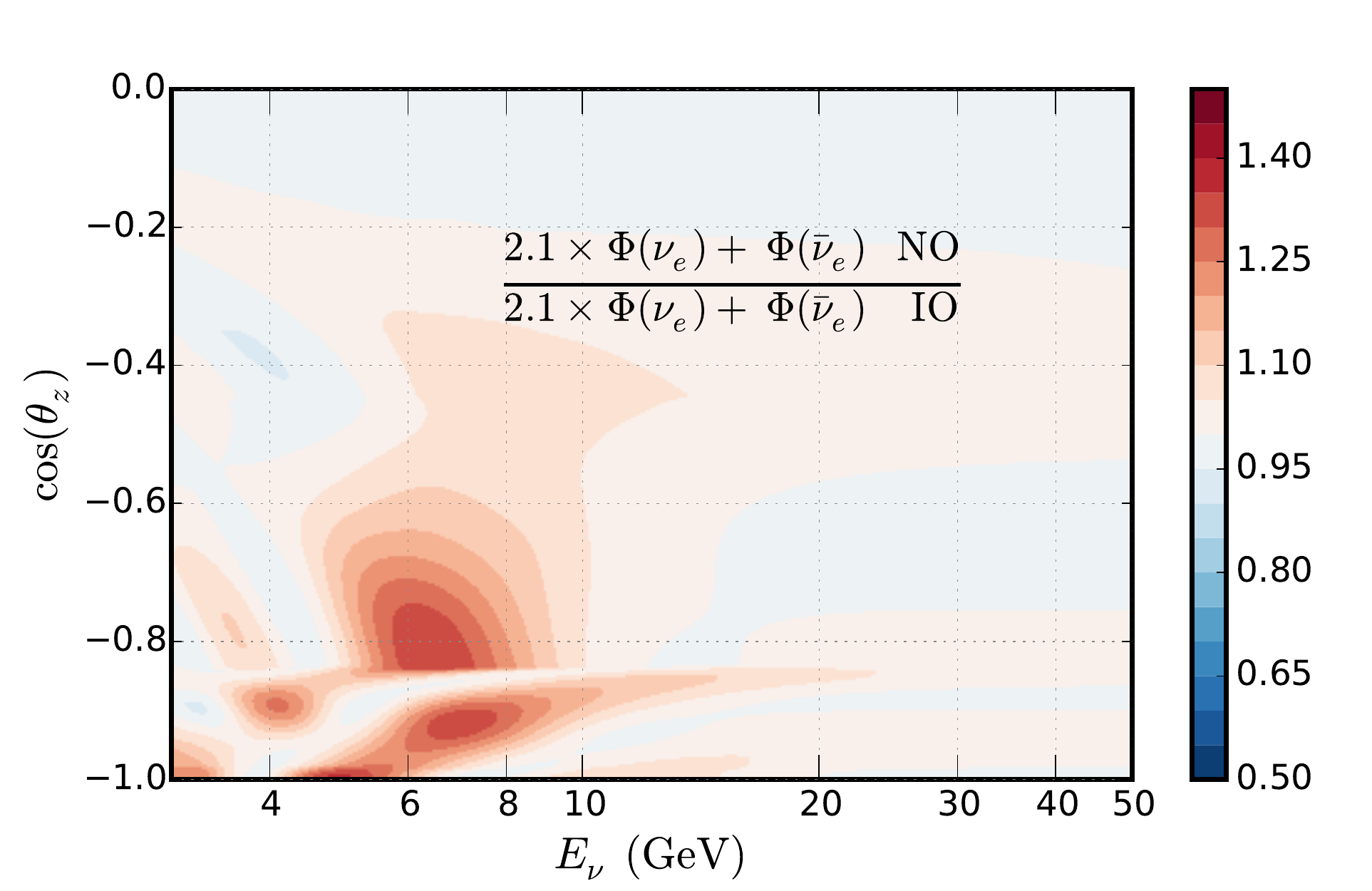}
  \caption{Expected interaction rate of electron neutrinos and antineutrinos predicted by a NO over the rate predicted assuming an IO. Using the oscillation parameters in \cite{Gonzalez-Garcia:2014bfa}. Because of the flux ratio $\nu_\mu/\bar{\nu}_\mu$ and the cross section difference, estimated to be 2.1 times larger for neutrinos than antineutrinos, more electron neutrino interactions are expected for a NO.}
  \label{fig:osc_prob3}
\end{figure}

The NMO can only be accessed with neutrinos below 15\,GeV, where matter induced resonances occur either for neutrinos or antineutrinos. The survival probability of muon (anti)neutrinos, the main component of atmospheric neutrinos, is modified by matter effects by about 20\%. As will be discussed in Section~\ref{sec:vlvnt}, VLVNT cannot separate neutrinos from antineutrinos event-wise, and instead rely on the $\nu / \bar{\nu}$ flux ratio and the difference in cross sections to identify whether oscillation probabilities of neutrinos or antineutrinos are modified by matter effects.

An interesting feature introduced by matter effects is that, instead of oscillating fully into $\nu_\tau$, muon neutrinos also change into $\nu_e$. Transitions of these type are almost symmetric between the two flavors (see Fig.~\ref{fig:osc_prob2}), but since the flux of $\nu_\mu$ is several times that of $\nu_e$ at the energy and zenith angle of interest (see Fig.~\ref{fig:atmnu_ratios}), the net effect is a significant excess of electron neutrinos with respect to the original $\nu_e$ flux. In the NO, the $\nu_e$ flux is enhanced, while for an IO the enhancement is realized for $\bar{\nu}_e$. Because of the initial $\nu_\mu/\bar{\nu}_\mu$ flux ratio and the differences in the $\nu_e/\bar{\nu}_e$ cross sections, different orderings result in a different number of detected events. Figure \ref{fig:osc_prob3} shows the ratio between expected interaction rates of $\nu_e + \bar{\nu}_e$ for normal and inverted orderings, including all of the oscillation channels. A factor of 2.1 is applied to neutrinos to account for the difference in cross sections. The normal mass ordering predicts up to 30\% more events in the region $E_\nu = [5, 8]$\,GeV and $\cos\theta_z = [-0.9, -0.5]$. Measurements of the flux of atmospheric electron neutrinos thus provide suitable data for determining the NMO.

The VLVNTs currently in operation are presented in details in the next section. With an energy threshold of about 15\,GeV they operate in the saturated regime. They can measure muon neutrino disappearance as well as tau neutrino appearance, and thus $\theta_{23}$ and $|\Delta m^2_{31}|$. Measuring the sign of $\Delta m^2_{31}$, on the other hand, requires measuring differences in oscillation probabilities below this threshold (see Figs.~\ref{fig:osc_prob1} and \ref{fig:osc_prob2}). This is the main goal of the next-generation detectors discussed in Section~\ref{sec:future}.
\section{Very large volume neutrino telescopes}
\label{sec:vlvnt}

A generic VLVNT is a three-dimensional array of photo-sensors detecting the Cherenkov light of charged particles produced after a neutrino interaction. The secondaries of neutrino interactions above a few GeV produce enough light so that they can be observed by sensors several meters apart. The spacing between the optical sensors define the energy threshold of VLVNTs, which is approximately 15\,GeV in currently operating detectors.

\subsection{VLVNTs in operation}

The optical sensors of VLVNTs are deployed at depths of 1\,km or more, in an optically transparent, naturally occurring medium. Sensors are laid out in lines or strings that are operationally independent. The spacing between sensors is uneven, being considerably larger in the $x-y$ plane (in between lines/strings) than in the $z$ plane. The sensors also have a preferred acceptance for light coming from below, although this might change for future detectors.

The neutrino telescopes currently in operation are IceCube in Antarctica \cite{icecube}, ANTARES in the Mediterranean Sea \cite{Collaboration:2011nsa}, and the prototype of the Gigaton Volume Detector in Lake Baikal \cite{baikal}. Both ANTARES and IceCube have published studies of neutrino oscillations, and are therefore the only ones discussed in this review.

\subsubsection{Detector design and layout}

\begin{figure}
        \centering
        \begin{subfloat}
                \centering
                \includegraphics[clip, trim=40 0 0 0, width=0.64\columnwidth]{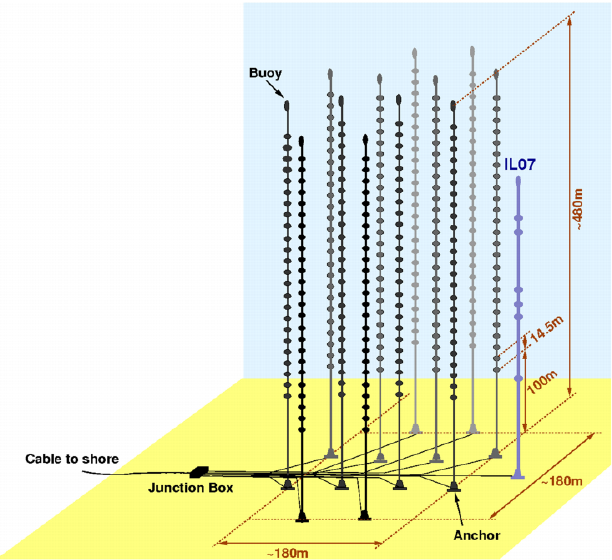}
        \end{subfloat}
        \begin{subfloat}
                \centering
                \includegraphics[width=0.7\columnwidth]{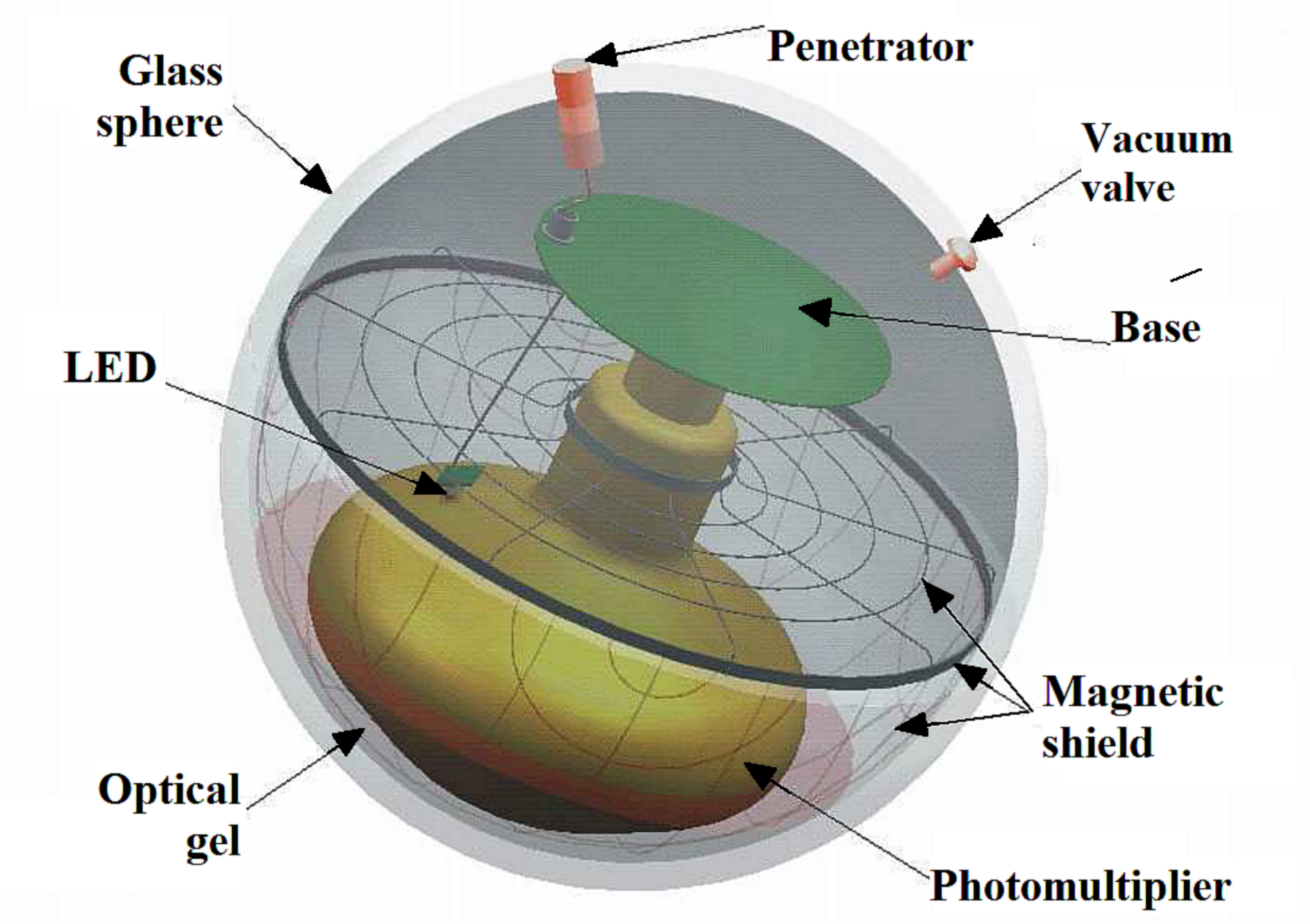}
        \end{subfloat}
        \caption{The ANTARES detector configuration (top drawing): the 12 detection lines are connected to a single junction box providing power and transferring all data recorded by the OMs to the shore station through a main electro-optical cable. The bottom sketch shows the OM and the components it houses, including a 10" photomultiplier tube.}
        \label{fig:antares_diagram}
\end{figure}

ANTARES is located between depths of 2025-2475\,m, 20\,km away from Toulon (French Riviera), in the Mediterranean Sea \cite{Collaboration:2011nsa}. It comprises 885 optical modules (OMs) \cite{ANTARES_OM}, distributed along 12 flexible lines. OMs are grouped in triplets, with 25 triplets per line. The distance between triplets is 14.5\,m, and the separation between lines ranges from 60 to 70\,m, as sketched in Fig.~\ref{fig:antares_diagram}. Acoustic devices, tiltmeters and compasses are used to monitor the shape of the detector, which is influenced by sea currents. 

\begin{figure}
        \centering
        \begin{subfloat}
                \centering
                \includegraphics[width=0.75\columnwidth]{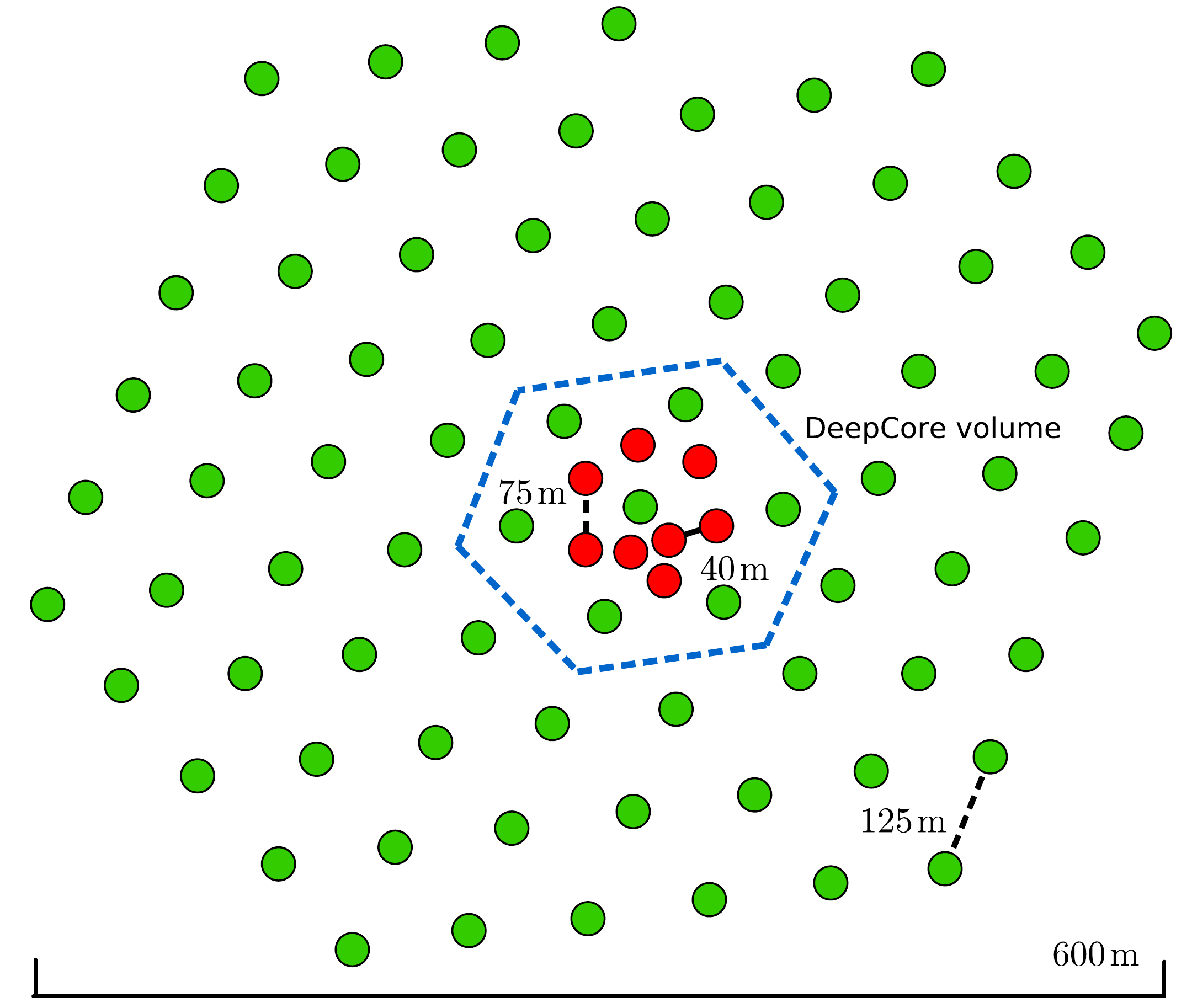}
        \end{subfloat}
        \begin{subfloat}
                \centering
                \includegraphics[width=0.75\columnwidth]{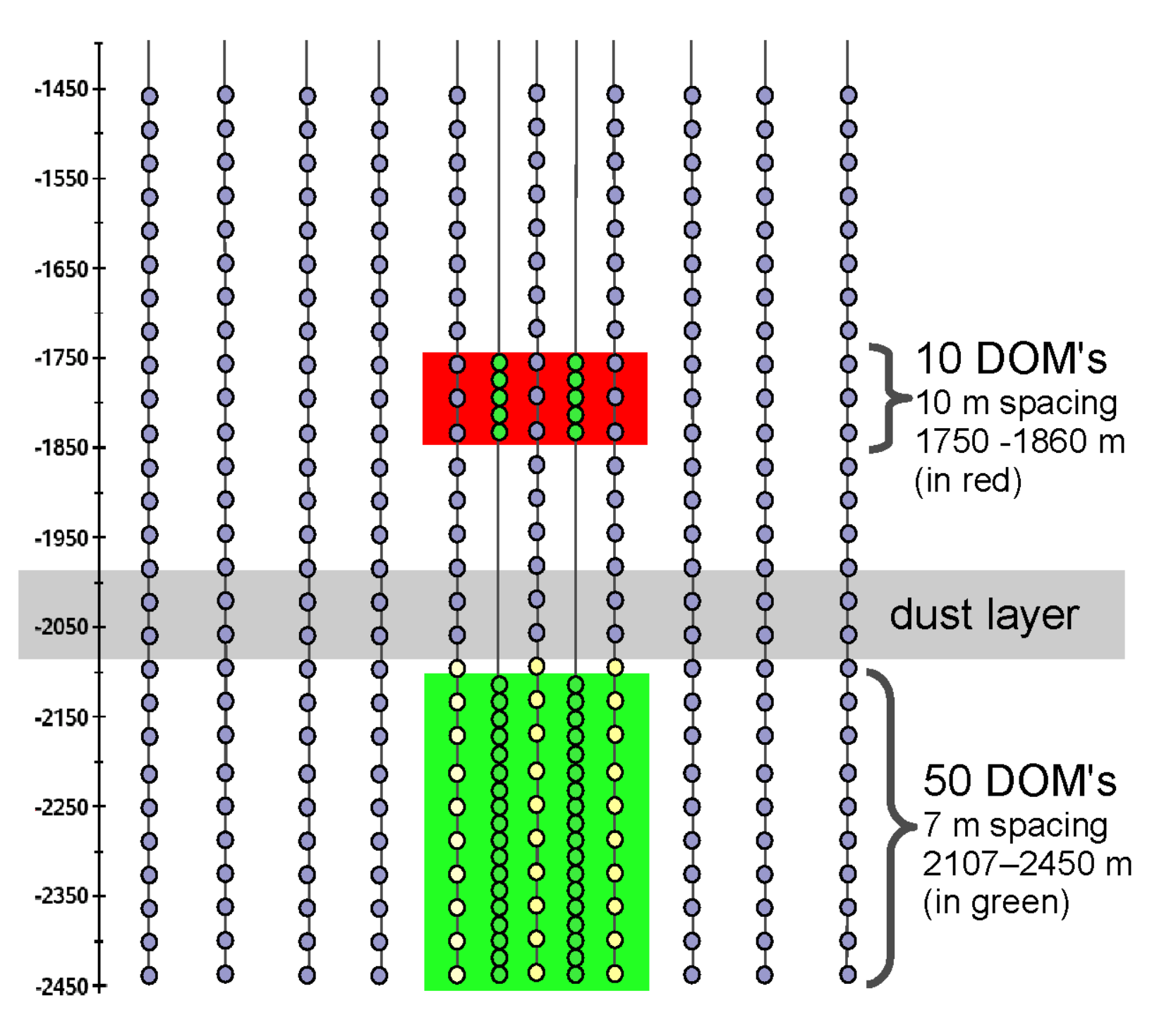}
        \end{subfloat}
        \caption{IceCube. Top and side schematic projections of the detector. The DeepCore volume used for analysis is highlighted in both figures.}
        \label{fig:icecube_diagram}
\end{figure}

IceCube is located at depths between 1450-2450\,m at the geographic South Pole \cite{icecube}. The in-ice part of IceCube consists of 5160 downward-facing digital optical modules (DOMs) \cite{icecube_daq}. The detector has 86 strings, each holding 60 DOMs. Of these, 78 strings are arranged in a hexagonal grid with a typical distance of 125~m (horizontal spacing) and 17~m (vertical spacing) between DOMs. A sketch of the detector layout is shown in Fig.~\ref{fig:icecube_diagram}.

The lower center region of IceCube, from 1760\,m down to 2450\,m, houses DeepCore \cite{dcdesign}, a region of denser instrumentation (7~m DOM vertical spacing), where eight strings are separated by $40-70$\,m. Some 50\,\% of the PMTs in this region have 35\,\% higher quantum efficiency than the standard IceCube PMTs. The DeepCore fiducial volume used for data analysis is defined by a cylinder with a height of 350\,m and a radius of approximately 150\,m that starts below a dust layer, where the light transparency is reduced, as shown in Fig.~\ref{fig:icecube_diagram}. This volume, which corresponds to roughly 2.5 times that of ANTARES, encloses about 550 DOMs with reduced spacing and results in a threshold for detection and reconstruction of neutrinos of about 15 GeV.

The optical modules of both IceCube and ANTARES are glass spheres enclosing a ten-inch PMT, optical coupling gel, and a $\mu$-metal cage for magnetic shielding. The IceCube OM digitizes the waveforms detected by the PMT inside the module before transmission \cite{icecube_daq}, while the ANTARES OM keeps the readout to a minimum and only transmits the time and amplitude of a signal above threshold \cite{ANTARES_electronics}. ANTARES optical modules have a baseline noise rate of 70\,kHz at single photon level \cite{ANTARES_noise}, while for IceCube (DeepCore) OMs the noise is 0.45\,kHz (0.65\,kHz) \cite{Abbasi:2010vc}.

\subsubsection{Optical medium and calibration}

The optical properties of the medium affect the time of arrival and the number of detected Cherenkov photons. At the ANTARES site (salt water) the absorption length, which is 60\,m for blue light ($\lambda\simeq$\,470\,nm) and 26\,m for UV light ($\lambda\simeq$\,375\,nm), reduces the number of photons observed. The effective scattering length, which is 256\,m for blue light and 122\,m for UV light, is considerably larger than the spacing between sensors \cite{ANTARES_light}. In the clear ice in which DeepCore is located the absorption length of UV light ($\lambda\simeq$\,400\,nm) is of the order of 200\,m, which is larger than the spacing between sensors. The effective scattering length in the deep Antarctic ice is approximately 50\,m, comparable to the string distance of DeepCore, thus significantly modifying the expected time of arrival of photons \cite{ice_mie, ice_lea}.

Water offers the advantage of being a homogeneous medium. Nonetheless, sea currents can deviate the detector lines so the position of the lines needs to be monitored constantly. This is achieved by combining acoustic triangulations with tilt and compass measurements yielding a precision better than 10~cm, which does not affect the angular resolution \cite{antares_amadeus}. High sea currents can also trigger bioluminescence bursts that must be accounted for in the optical background simulation, in addition to the stable optical noise arising for $^{40}$K decays. The latter can be used for determining the absolute detection efficiency of the optical modules. 

In ice, the positions of the optical modules are fixed and known to within a few cm. Noise levels are constant and a hundred times lower than in salt water after the detector has stabilized. A disadvantage of using ice is that the medium is not homogeneous and its structure has to be modeled. This is particularly challenging in the immediate surroundings of the optical modules. Columns of the original glacier are melted to deploy the instrumentation. The refreezing process leaves behind clear ice near to the boundaries of the hole, and a cylinder of ice of about 10\,cm in diameter with a high concentration of bubbles towards the center of the column. These changes in ice properties modify the DOM angular acceptance measured in the laboratory. Future detectors in ice will consider the possibility of de-gassing the water to avoid trapping air bubbles inside the hole ice, and with that reduce the impact of the medium.

The absolute optical efficiency of the optical modules as well as their angular acceptance must be determined in situ after deployment. ANTARES and IceCube use both controlled light sources and minimum ionizing muons to calibrate the efficiency and timing accuracy of their optical modules \cite{antares_timecal1,antares_timecal2, icecube_energy}. Relative arrival times are known with a precision better than 3\,ns and 1.5\,ns for IceCube \cite{icecube_daq} and ANTARES, respectively.

\subsection{Neutrino interactions}
\label{sec:nu_interactions}
\begin{figure}[bth!]
        \centering
        \begin{subfloat}
                \centering
                \includegraphics[width=0.95\columnwidth]{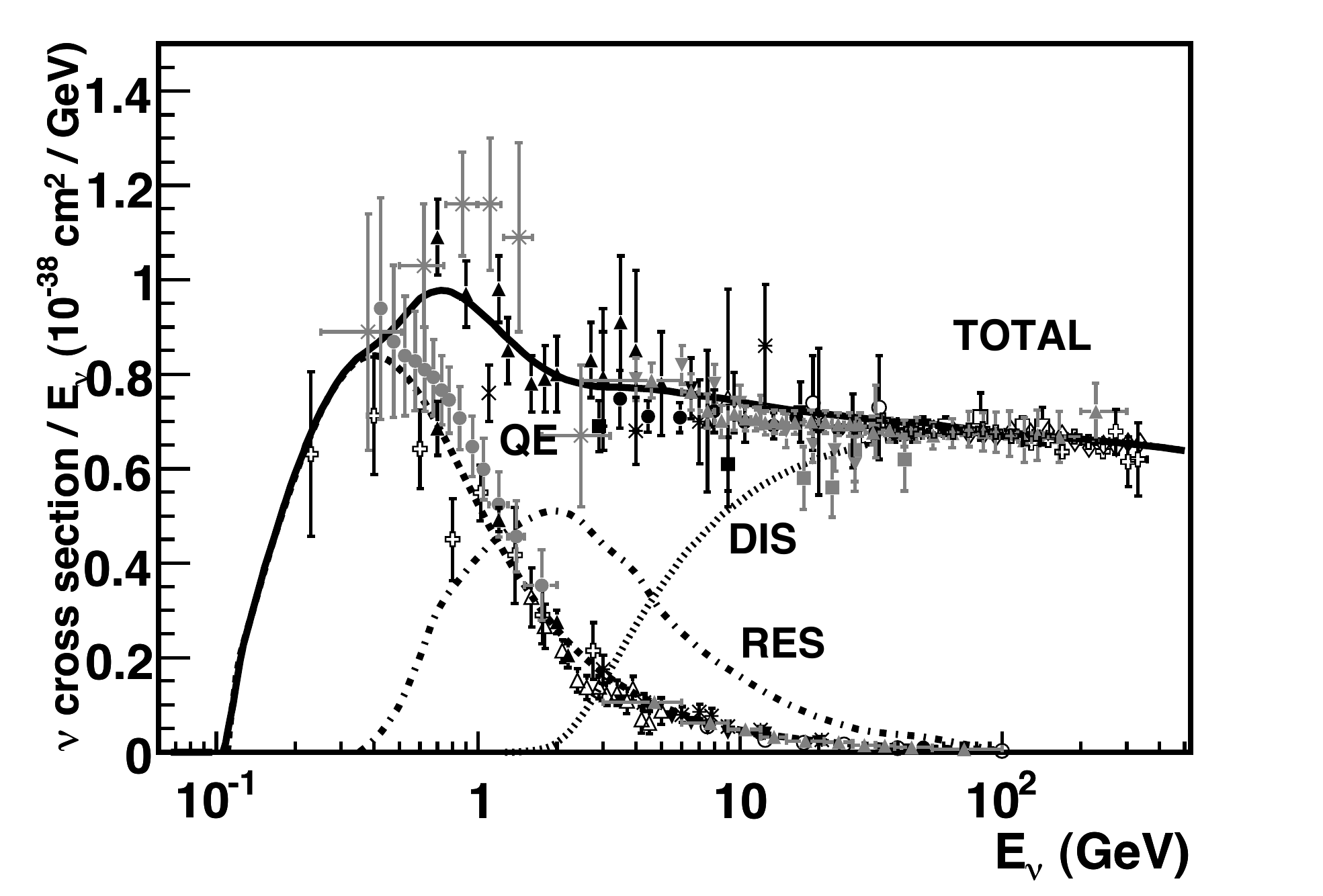}
        \end{subfloat}
        \begin{subfloat}
                \centering
                \includegraphics[width=0.95\columnwidth]{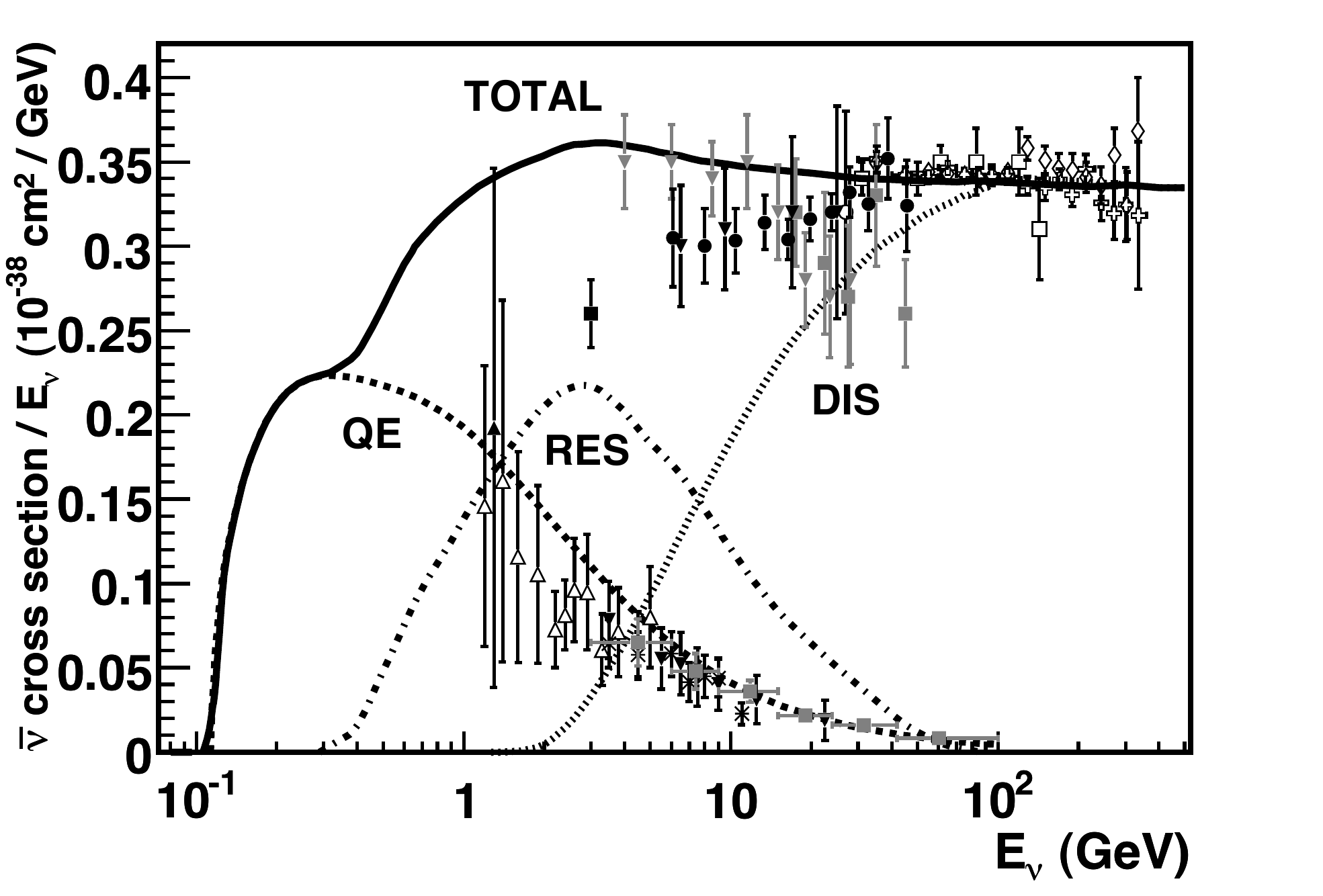}
        \end{subfloat}
        \caption{Collection of existing muon neutrino (top) and antineutrino (bottom) charged current cross section measurements and predictions as a function of neutrino energy (see \cite{Formaggio:2013kya} for details on the experiments contributing to the data points and \cite{Casper:2002sd} for a description of the model used). The contributing processes in this energy region include quasi-elastic (QE) scattering, resonance production (RES), and deep inelastic scattering (DIS). Taken from \cite{Formaggio:2013kya}.}
        \label{fig:neutrino_xs}
\end{figure}

The dominant neutrino interaction for most of the energy range that VLVNTs can access is neutrino-nucleon deep inelastic scattering (DIS), with other processes being only a sub-dominant contribution. Nonetheless, below 15\,GeV, the region of interest to search for matter effects in neutrino oscillations and the NMO, quasi-elastic scattering and production of resonances compete with DIS processes. Figure\,\ref{fig:neutrino_xs} shows a calculation of the competing $\nu N$ cross sections around the GeV region, together with the data available.

Most of the knowledge of neutrino-nucleon cross sections between $1-15$\,GeV comes from bubble chambers or spark chamber detectors which collected comparatively small data samples. Thus, the constraints on the models that describe them are rather weak \cite{Formaggio:2013kya}. The uncertainty with the largest impact on the neutrino cross sections for quasi-elastic and resonant interactions, which changes them by up to 40\%, is the value of the axial mass that effectively describes the nucleon form factor and has an estimated error of 15\%-25\% \cite{Kajita:2005hj,Formaggio:2013kya}. DIS interactions in the crossover region have a small momentum transfer. Non-perturbative QCD calculations are required \cite{Bodek:2002ps}, and the estimated errors are as well of the order of 20\% \cite{sarkar}.

Deep inelastic scattering accounts for 90\% or more of the total cross section of neutrinos and antineutrinos above an energy of roughly 12\,GeV, as shown in Fig.\,\ref{fig:neutrino_xs}. DIS in the perturbative regime is comparatively better understood than the processes discussed so far, with uncertainties coming mainly from the determination of the parton distribution functions (PDFs) of the nucleons. The uncertainties on the PDFs change the total cross section by 5\% or less \cite{sarkar}.

At these energies the neutrino-nucleon DIS charged current (CC) cross section is quasi-independent of the inelasticity $y$ ($y=1-E_\mathrm{lepton} / E_\nu$) of the interaction, while for antineutrinos the cross section is accompanied by a factor $(1-y^2)$, which suppresses kinematic configurations where the hadronic part of the interaction takes most of the energy. The inelasticity dependence makes the total $\bar{\nu}N$ cross section about one half of that of $\nu N$. PDF uncertainties impact the inelasticity dependent cross section by only a few percent \cite{Connolly:2011vc}.

While the neutrino-nucleon DIS CC cross sections for $\nu_e$ and $\nu_\mu$ are equal, the $\nu_\tau N$ one is suppressed due to the mass of the tau lepton. It is only at $E_\nu \sim 40$\,GeV that the cross section reaches half of the value of the other neutrino flavors \cite{donut}.

In neutral current interactions (NC) one or several hadrons are produced, initiating a hadronic shower. In charged current (CC) interactions a hadronic shower is also present, but now the neutrino transforms into a charged lepton. Electrons and taus also initiate a shower of particles after they are produced\footnote{The tau lepton has a 17\% probability to decay into a muon. However, due to energy losses and other particles involved in the processes, muons from tau decays with a range larger than a few meters are uncommon.}. Muons, on the other hand, travel practically undisturbed and lose energy at a quasi-constant rate. For muons passing through water $dE/dx \sim 0.25 $\,GeV/m up to a few hundred GeV \cite{pdg2014}. Once they travel distances comparable to the detector spacing they can be identified, and charged current $\nu_\mu$ interactions can be tagged.

\subsection{Event reconstruction}

Neutrino interactions are reconstructed using the number of photons recorded by the optical module (or time over threshold), as well as the time at which they arrive. For the energy range under consideration, the most general hypothesis is an interaction which produces a hadronic shower (all interaction types), an electromagnetic shower ($\nu_e$ CC) or a long range muon ($\nu_\mu$ CC). The direction in which these particles are produced is reconstructed from the arrival times of the emitted photons. The Cherenkov light of muons is produced almost perfectly in a cone. The light coming from the cascade is also beamed in the Cherenkov angle, but the smearing due to multiple particle contributions to it is larger, which degrades the achievable precision of directional reconstructions. This smearing effect is stronger for hadronic showers.

The energy reconstruction of showers is primarily given by the number of photons detected from a given interaction, and its accuracy depends mainly on the reconstructed position of the interaction vertex. To estimate the energy an assumption has to be made on whether the shower is hadronic or electromagnetic. The energy of muons can be estimated by the observed range in the detector.

In principle it is possible to fit the directions of both cascade and track components in an interaction. The sparse instrumentations of the detectors, however, make it challenging. In the simplest approach, tracks and cascades are assumed to be collinear.

\subsection{Simulation tools}

\begin{table*}[bth!]
\centering
\begin{Tabular}[1.3]{c|c|c|c|c|c}
%\hline
  & \multirow{2}{*}{Parameter} & \multicolumn{2}{c|}{VLVNT}& \multirow{2}{*}{SK} & \multirow{2}{*}{MINOS, T2K, NOvA} \\ 
&& ANTARES& DeepCore& & \\
\hline \hline
\multirow{5}{*}{\begin{sideways}Detector (far)\end{sideways}} &Instrumentation density (m$^{-3}$)   & $9.1\times 10^{-5}$\,OMs & $2.3 \times 10^{-5}$\,DOMs & 0.2\,OMs& 15 channels\\
&Detection principle       & \multicolumn{2}{c|}{Cherenkov light over tens of meters} & Cherenkov rings & Trackers/calorimeters\\
&$E_\nu$ resolution         &   50\% $\pm$ 22\% & 25\% at 20 GeV & 3\% at 1 GeV & 10-15\% at 10 GeV \\
&$\theta_\nu$ resolution    & 3$^\circ$ at 20\,GeV & 8 $^\circ$ at 20\,GeV & 2-3$^\circ$ & --\\
&Particle ID capabilities  & \multicolumn{2}{c|}{Muon/no muon in interaction} & $e,\,\mu,\,\pi$ (rings) & Individual particles, charge\\
\hline
\multirow{7}{*}{\begin{sideways}Neutrino flux\end{sideways}}&Source of neutrinos       & \multicolumn{3}{c|}{Atmosphere: mix of $\nu_e,\,\bar{\nu}_e,\,\nu_\mu,\,\bar{\nu}_\mu$} & Accelerator: $\nu_\mu/\bar{\nu}_\mu$ modes\\
&Baseline                  & \multicolumn{3}{c|}{10-12700\,km} & 300-800\,km\\
&Flux determination        & \multicolumn{2}{c}{Atm. $\nu$ models, self fit} &  + top/down ratios & Near/Far detector\\
&Energy range              & \multicolumn{2}{c|}{10$-$100\,GeV} & few MeV$-$few GeV & few GeV \\
&Main interaction channel  & \multicolumn{2}{c|}{DIS} & QE & QE, RES, COH, DIS\\
&$\nu$ events expected with osc.        &  530 &  1800 & 2000 & 30\,(T2K), 900\,(MINOS)\\
&and without osc. (per year)            & 660  & 2300 & 2300 & 120\,(T2K), 1050\,(MINOS) \\
\hline
\end{Tabular}
\caption{Qualitative comparison of experiments measuring the atmospheric neutrino oscillation parameters. The table is divided in detector and flux characteristics. Note that the far detector of T2K is Super-Kamiokande, but uses accelerator neutrinos. Detector performances taken from \cite{Collaboration:2011nsa, antares_osc, icecube_yanez, Wendell:2010md, T2K_latest, minos1, MINOS}. Expected neutrino events quoted from published results of $\nu_\mu$ disappearance at analysis level (note that for VLVNTs this number can vary significantly depending on the studied range in energy, zenith angle and topology). COH refers to coherent pion production. For details on the other interaction channels and energy ranges see Fig.~\ref{fig:neutrino_xs}.}\label{table:experimental}
\end{table*}

The measurement of diffuse fluxes in VLVNTs, such as the one required to determine oscillation parameters, rely fully on the correct modeling of the experimental set-up. Atmospheric muons, the leading source of background, are simulated in IceCube using full showers and parameterizations obtained from CORSIKA \cite{corsika}. ANTARES uses the MUPAGE program, which produces muons based on a parameterization tuned to MACRO data \cite{mupage}.

Neutrino interactions in IceCube are simulated using the GENIE package \cite{genie} ($E_\nu \leq 200$\,GeV) and NuGen/ANIS \cite{nugen} ($E_\nu \geq 50$\,GeV). Besides GENIE, ANTARES uses an in-house neutrino generator based on LEPTO \cite{lepto_mc} for the full energy range, with the PYTHIA package \cite{pythia} handling the hadronization processes. The neutrinos produced are then weighted to match the flux predictions of the Honda and/or Bartol groups \cite{honda2012, Barr:2004br}.

The propagation of short-ranged particles produced in the interaction is done, both in IceCube and ANTARES, using the Geant software \cite{geant4} as basis. Parameterizations of the light yield of these particles are produced by both experiments, and used to obtain the detector response to high energy hadrons, electrons and photons \cite{brunner_sim}, while low energy hadrons ($E \leq 30$\,GeV) are propagated individually. Muons are propagated using code optimized for simulation of long ranged leptons, namely MUSIC \cite{antares_music} and MUM \cite{antares_mum} in ANTARES, and MMC \cite{mmc} in IceCube.

The Cherenkov photons produced during the propagation of charged particles are individually traced through the ice in IceCube/DeepCore, while ANTARES uses expectation from tables. Low energy future projects (Section~\ref{sec:future}) foresee using only individual tracing to assure that the optical properties of the medium are included in detail. After the photons are propagated, the response of the optical module is recreated, and events from simulation and experimental data are treated equally.

\subsection{Large statistics vs. precise reconstruction}

The current knowledge on the oscillation parameters in the atmospheric sector comes from experiments which differ from VLVNTs substantially: they are Super-Kamiokande \cite{superk}, T2K \cite{t2kdesign}, MINOS \cite{MINOS}, which is no longer in operation, and the recently commissioned NOvA \cite{nova}. Table~\ref{table:experimental} contains a qualitative comparison of the detectors and neutrino sources used by these experiments and VLVNTs.

Super-Kamiokande, which also measures atmospheric neutrinos, has about twenty (ten) times the number of optical sensors as DeepCore (ANTARES), separated by a few cm, placed on a cylindrical tank with a diameter similar to the inter-string distance in ANTARES/DeepCore. Neutrinos are detected using the rings produced after the Cherenkov light of the charged products of the interaction hits the walls of the detector. Muons, electrons and pions can be identified by the differences in the ring pattern they produce. Because of its considerable smaller size and the steepness of the spectrum of atmospheric neutrinos, its operating energy is lower than that of VLVNTs.

Long baseline experiments, such as T2K, MINOS and NOvA, use neutrinos from particle accelerators and have near and far detectors. While T2K uses Super-Kamiokande as a far detector, MINOS and NOvA follow an experimental set-up where the far detector is smaller than Super-Kamiokande, but is more densely instrumented, can be magnetized, and observes the path of individual particles coming from a neutrino interaction. These set-ups benefit from their controlled neutrino source and detailed event reconstruction. Unlike the case of atmospheric neutrino experiments, long baseline experiments have a unique baseline and cover a narrow energy range, allowing for better precision but also limiting the $L/E$ region that they can access. It should also be noted that, as stated in Section~\ref{sec:nu_interactions}, the poor knowledge of neutrino interactions at energies of a few GeV introduces significant uncertainties in the data analysis of long baseline oscillation experiments.

VLVNTs have become competitive with accelerator based experiments thanks to the possibility of observing multiple combinations of baseline and energy ($L/E$), and with Super-Kamiokande because VLVNTs can collect large event samples and in an energy range where most events are DIS which can be modeled with high accuracy. The sparse instrumentation does not permit observation of small details of the interaction, but in the same way reduces the impact from uncertainties in the hadronization processes, one of the leading systematic uncertainties for MINOS \cite{MINOS_unc} and T2K \cite{T2K_latest}. Reconstruction accuracy and proper handling of systematic uncertainties are the most important points to consider for precision measurements with VLVNT. 

\section{Neutrino oscillation measurements from running VLVNTs}
\label{sec:results}
The ANTARES and IceCube collaborations have published measurements of oscillations studying the muon neutrino disappearance channel. Above 15\,GeV, where these detectors operate, muon neutrinos oscillate into tau neutrinos, following Eq.~\ref{eq:osc_musimple}. Signal neutrinos, i.e. $\nu_\mu$ interacting via CC with $E_\nu \sim 25$\,GeV, are typically recorded by a handful of optical modules both for ANTARES and IceCube's DeepCore. The events develop over a distance of order of 100\,m, and thus can be fully contained in both detectors.

The measurement of neutrino oscillations in VLVNTs follows a general strategy which begins with the reduction of the dominant sources of background, i.e. atmospheric muons and pure noise. Straight cuts are applied on variables of which the distribution for neutrinos differs from that of background sources. They generally aim for a neutrino purity higher than 95\%.

For the currently published results of both experiments, the presence of a muon in a neutrino interaction is required for an event to be selected for analysis. The analyses are done by comparing the histograms of data and simulation as a function of the reconstructed variable(s) used. The simulation is modified by the physics parameters of interest, $\theta_{23}$ and $\Delta m^2_{32}$, and by nuisance parameters which absorb the systematic uncertainties involved in the measurement. Errors are derived from a scan of the likelihood landscape, and/or directly using a $\chi^2$ approximation.

The results of ANTARES and IceCube that have been made public until now use only events coming below the horizon. ANTARES removes the down-going region because it is dominated by atmospheric muons. IceCube uses the instrumentation outside DeepCore to veto atmospheric muons, nevertheless the contribution of these muons in the down-going region is still significant, so the region is also removed from analysis. This situation is different for Super-Kamiokande, where events from the entire zenith range are used in oscillation studies and top-down ratios are used to reduce uncertainties. Ongoing studies within IceCube are exploring the possibility of using neutrinos coming from above the horizon in future results \cite{jp_vlvnt}.

\subsection{First measurements of oscillations from ANTARES}
The ANTARES collaboration presented the first results on the study of neutrino oscillations from a VLVNTs \cite{antares_osc}. The analysis relied on the muon track reconstruction described in~\cite{antares_bbfit}, which fits the depth at which the Cherenkov cone of light arrives at the OMs as a function of time. This corresponds to a hyperbola of which the orientation of the asymptotes depends on the zenith angle. An algorithm that searches for these patterns, without assuming any knowledge on the arrival angle of the emitter, was implemented. The algorithm is capable of rejecting noise hits and keeping events down to energies of 20\,GeV ($R_\mu = $100\,m) with photons in a single line, and 50\,GeV ($R_\mu=250$\,m) in multiple lines. Mis-reconstructed muons that appear up-going are removed by selecting only events which have a good fit quality. This cut also effectively reduces the contribution of NC interactions from all flavors, and $\nu_e$ CC interactions.

The median zenith angle resolution with respect to the neutrino direction of single-line events is 3.0$^\circ$, and it reduces to 0.8$^\circ$ for multi-line events. The energy of the neutrino is estimated solely by the muon range, resulting in a lower limit to the neutrino energy, where $E_\mathrm{reco} = (50\% \pm 22\%) E_\nu$.

The analysis is done by comparing data and simulation as a function of $E_\mathrm{reco} / \cos\theta_\mathrm{reco}$ by means of a $\chi^2$, combining single- and multi-line selections. Only events below the horizon ($\cos\theta_\mathrm{reco} < -0.15$) are considered. Systematic uncertainties are implemented using two normalization coefficients, for single- and multi-line events, as pull factors in the $\chi^2$ following the method presented in \cite{pulled_chi2}. These factors absorb the effects of changes in the average quantum efficiency ($\pm 10$\%), optical properties of sea water ($\pm 10$\%), the spectral index of atmospheric neutrinos ($\pm 0.03$), and disagreements between data and simulation during the selection (varying cut values). The overall normalization of the $\nu_\mu$ flux and detector efficiency are left unconstrained.

\begin{figure}[bt!]
 \includegraphics[width=0.99\columnwidth]{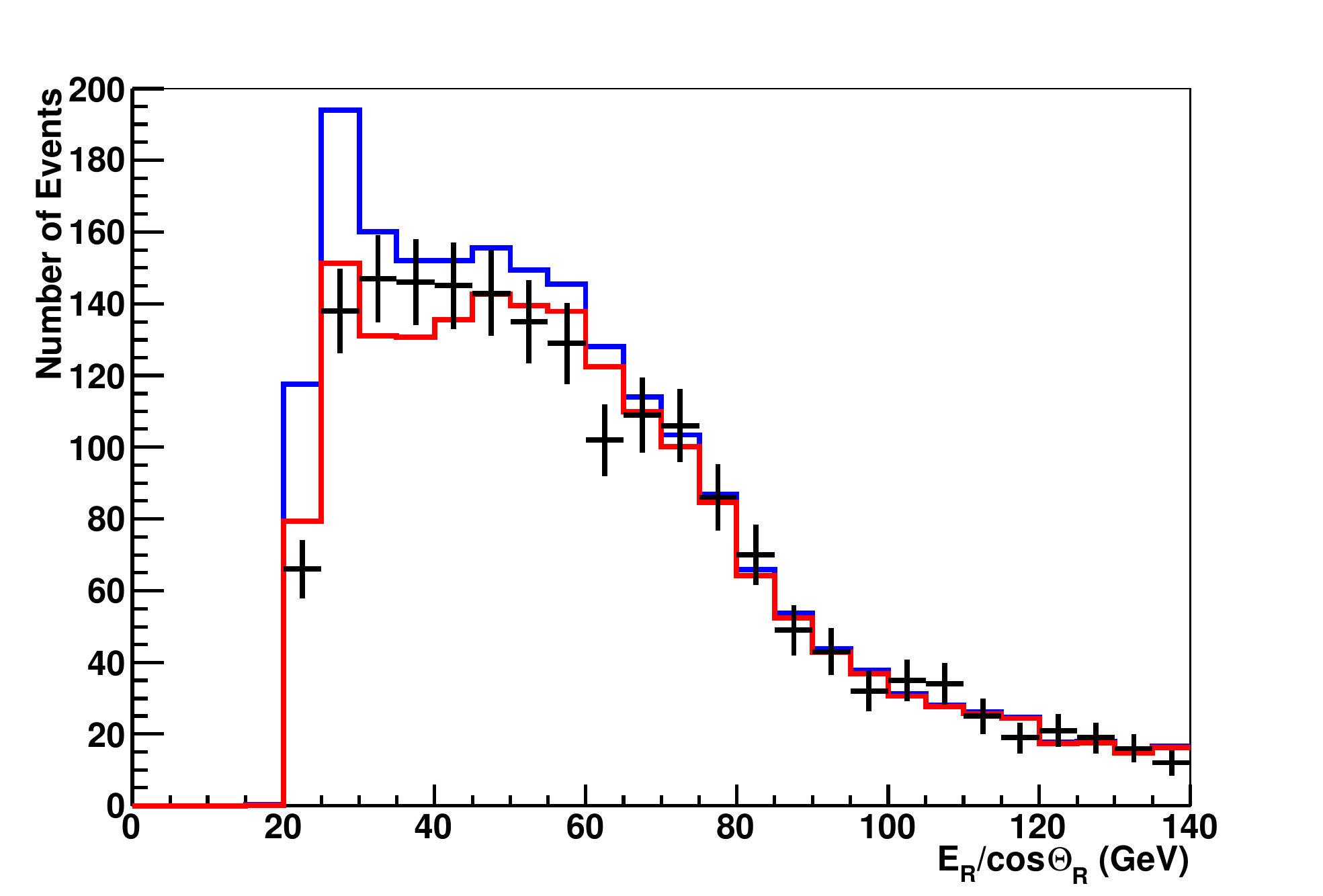}
 \caption{\label{fig:antares_results}Distribution of $E_\mathrm{reco} / \cos\theta_\mathrm{reco}$ for events selected in the oscillation analysis of ANTARES. Data are shown in black, simulation without oscillations in blue, and simulation with the fit parameters is given in red. From \cite{antares_osc}.}
\end{figure}

The data analyzed were taken between March 2007 and December 2010, corresponding to a detector live time of 863 days. A total of 2126 neutrino candidates were selected. The measured oscillation parameters, which were found to be compatible with the world's average, are indicated in Fig~\ref{fig:osc_compilation}. Data and simulation were in good agreement, as it can be seen in Fig.~\ref{fig:antares_results}, which results in a $\chi^2$/NDF = 17.1/21. The case of no oscillations could be rejected at the 3$\sigma$ confidence level. The ANTARES collaboration will proceed to an updated analysis of this kind with the full data sample collected until the end of the data taking, circa 2017.

\subsection{First measurements from IceCube DeepCore}
To this date, IceCube has reported results of four neutrino oscillation analyses of the low-energy DeepCore data. The selection, reconstruction and analysis methods have been refined in each step. The low-energy data for all studies comes from the DeepCore filter and trigger \cite{dcdesign}. The main source of background at this stage are triggers due to sensor self-noise and atmospheric muons. The instrumentation outside the fiducial volume of DeepCore (see Sec.~II-A and Fig.~\ref{fig:icecube_diagram}) is used to tag atmospheric muons. Low energy neutrino interactions are required to start within the DeepCore fiducial volume, while no requirement is imposed for full containment. 

Systematic uncertainties are accounted for using additional parameters which modify the expected number of events. An energy dependent term ($E^{-\gamma},\,\gamma \pm 0.05$) and a free overall normalization absorb total cross section uncertainties, and the uncertainties on the spectral index of the neutrino flux. The electron neutrino flux is varied by $\pm20$\% around the predicted value. The cosmic ray models which predict the cosmic muon contamination are varied to obtain a robust estimate. The effects of changing the optical description of the pristine ice, as well as the re-frozen ice around the DOMs, are studied by producing multiple simulation sets.

\begin{figure}[hbt]
 \includegraphics[width=0.95\columnwidth]{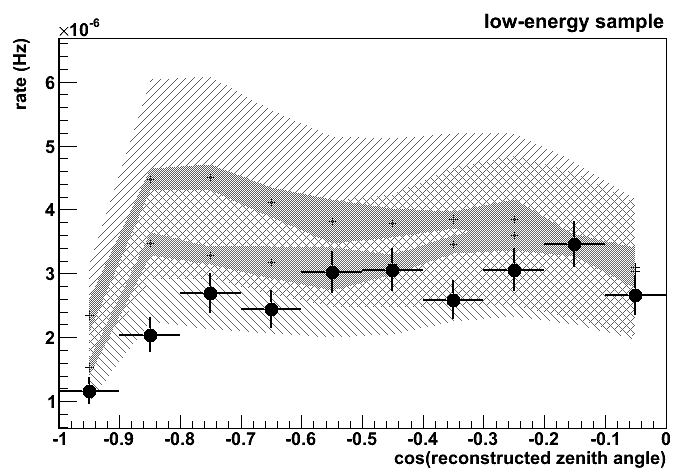}
 \caption{\label{fig:icecube_gross}Data and simulation expectation at world average oscillation parameters (in black) and the case of no oscillations (in red) for the low energy sample of IceCube's IC79-A analysis. Systematic uncertainties are split into a fully correlated part (hatched bands) and uncorrelated part (shaded bands). From \cite{icecube_gross}.}
\end{figure}

The initial three oscillation studies from DeepCore, presented first herein, were restricted to a single year of detector live time. Two used a partial configuration (IC79, two DeepCore strings missing) and one used the full detector (IC86). The first analysis \cite{icecube_gross}, from here on IC79-A, used a DeepCore low-energy sample where the effect of oscillations is expected ($E_\nu < 100$\,GeV, 719 events) and an IceCube high-energy sample, where oscillations play no role, to constrain flux and detection uncertainties ($E_\nu \simeq 1$\,TeV, 39638 events). The measurement was done by analyzing the distribution of events as a function of zenith angle in the low-energy sample (see Fig~\ref{fig:icecube_gross}). The zenith angle of both samples was estimated using the muon track reconstruction described in \cite{amanda_reco}. Atmospheric muons were mainly removed by reconstructing all events as up-going, and making cuts on parameters related to the quality of the reconstruction (without muon tagging).

\begin{figure}[bt]
 \includegraphics[width=0.9\columnwidth]{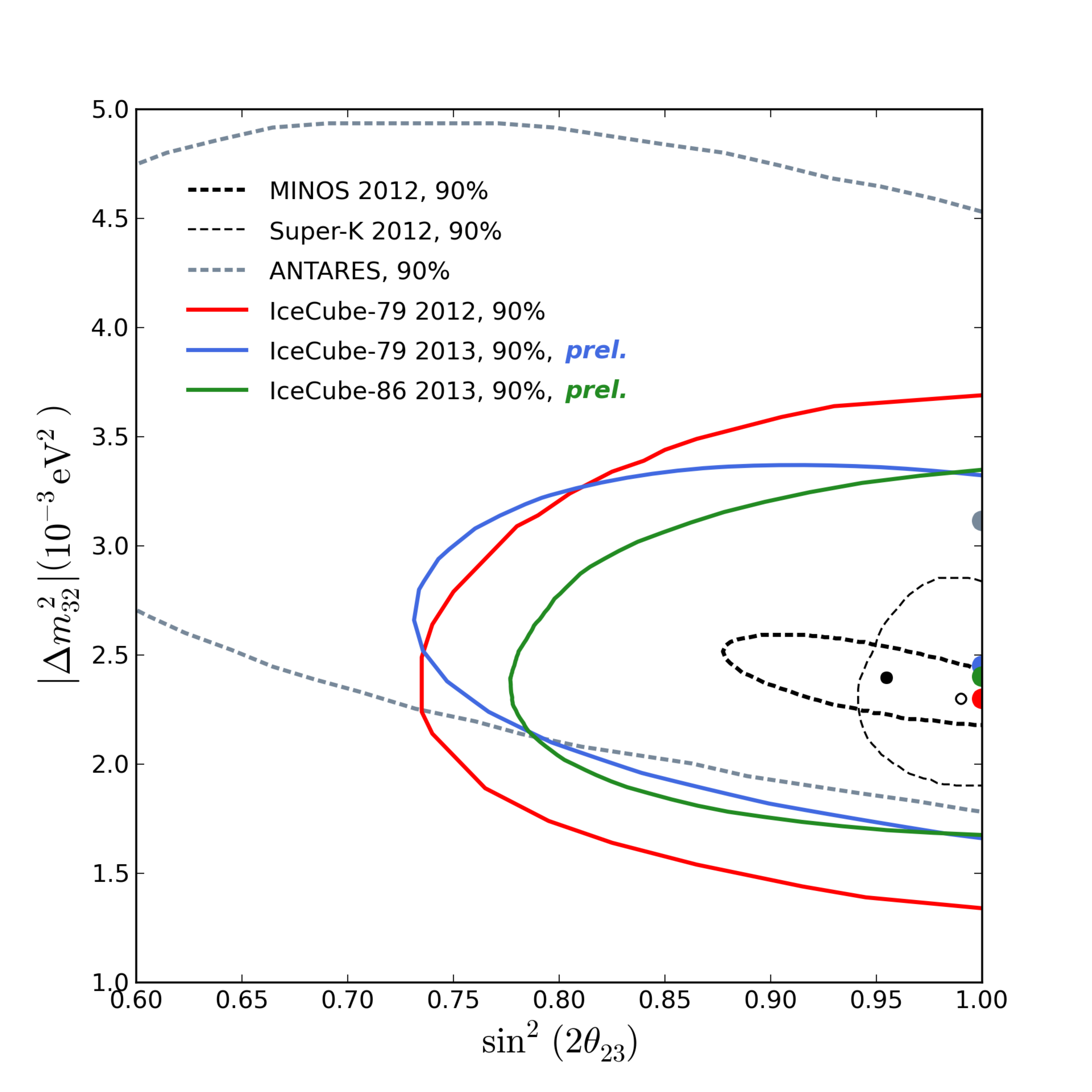}
 \caption{\label{fig:osc_compilation} 90\% C.L. contours from ANTARES \cite{antares_osc} and IceCube's single year measurements \cite{icecube_gross, icecube_icrc_euler, icecube_icrc_yanez}, compared to earlier measurements by MINOS \cite{minos1}, T2K \cite{T2K_old_result} and Super-Kamiokande \cite{sk_old_results}. Taken from \cite{brunner_review}.}
\end{figure}

The data were analyzed using a $\chi^2$ optimization with pulls, also following the method in \cite{pulled_chi2}. The results obtained for the atmospheric oscillation parameters were compatible with contemporary global fits \cite{GonzalezGarcia:2012sz}, although the errors were a factor 4 to 9 larger (see Fig.\,\ref{fig:osc_compilation}).

Two subsequent analyses of the data, from here on IC79-B and IC86-A, created new event selections based on the rejection of atmospheric muons by using the veto, separating the background rejection from the reconstruction of events \cite{icecube_icrc_euler, icecube_icrc_yanez}. In both cases only the low-energy DeepCore data were analyzed.

\begin{figure}[hbt]
 \includegraphics[width=1.\columnwidth]{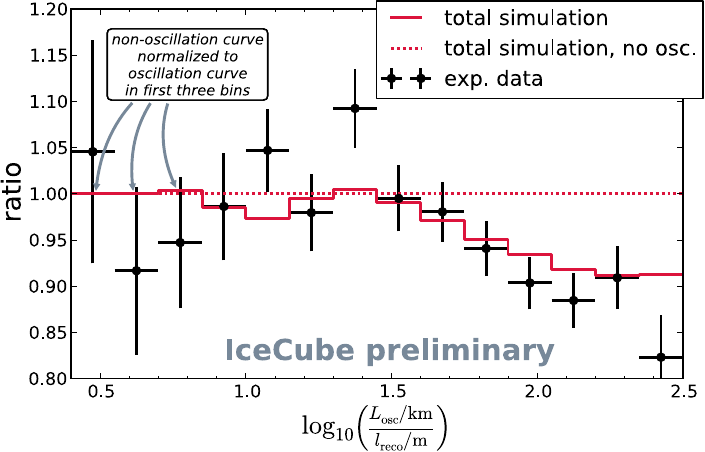}
 \caption{\label{fig:icecube_euler} Ratio of the distribution of oscillation length over reconstructed track length to the no-oscillation hypothesis from simulation in the IC79-B analysis. The best fit is also shown. From \cite{icecube_icrc_euler}.}
\end{figure}

The data used for IC79-B were acquired during the same period of time as for IC79-A, however, due to the change in the selection of events the final sample studied was a factor 10 larger. The zenith angle of events was reconstructed with a similar method as in IC79-A \cite{icecube_icrc_euler}. A second observable, the reconstructed muon range $L_\mathrm{muon}$ \cite{freco} was used as an energy proxy and the data were analyzed as a function of both observables. The ratio of events with respect to the no oscillation scenario, together with the best fit, are shown as a function of reconstructed $L_\mathrm{osc}/L_\mathrm{reco}$ in Fig.\,\ref{fig:icecube_euler}, where $L_\mathrm{osc}$ is the distance the neutrino traveled and $L_\mathrm{reco}$ is the reconstructed length of the muon produced in the interaction. The best fit and estimated errors of this method were similar to those of IC79-A.

\begin{figure*}[bht!]
 \includegraphics[trim = 5 0 10 0,width=1.\textwidth]{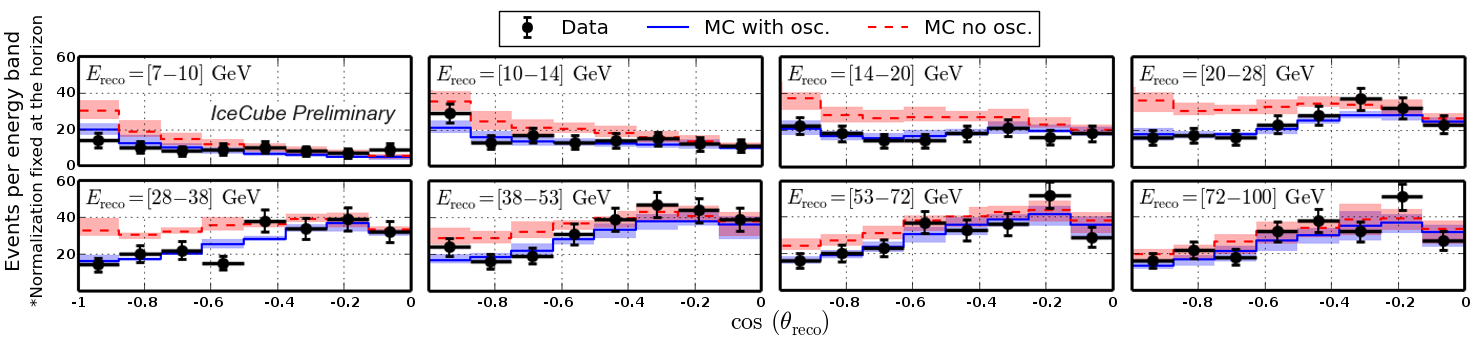}
 \caption{\label{fig:icecube_yanez1}Comparison between data and simulation for the two-dimensional histogram used in the IC86-A analysis of IceCube. The data are shown as a function of the zenith angle for the energy bins studied. Bands indicate the impact of the estimated systematic uncertainties. Figure taken from \cite{icecube_yanez_proc}.}
\end{figure*}

The first analysis of data from the full detector configuration \cite{icecube_icrc_yanez}, IC86-A, was performed using a selection of photons and event reconstruction based on the method published by ANTARES \cite{antares_bbfit}. The selection of photons was modified to remove multiply scattered photons instead of noise. Unscattered, or \textit{direct}, photons were identified by restricting their possible arrival times to those given by the hyperbolic pattern that Cherenkov light produces as a function of time as it crosses a string. About 70\% of the neutrino interactions which trigger the detector do not have a clear core of direct photons, and thus are removed. 

The \textit{direct} photons found are used to fit track and cascade hypotheses. The zenith angle from the track fit was used as an observable, and the ratio of the $\chi^2$ of the track and cascade fits was used to separate track-like from cascade-like events. An estimator of the total energy of the neutrino was also implemented, which takes the muon range estimator from IC79-A and also fits a hadronic cascade at the vertex.

In IC79-B and IC86-A the data were analyzed using a likelihood optimization with nuisance parameters to account for systematic uncertainties. For IC86-A, uncertainties related to the detector were also included as nuisance parameters. Simulation sets with varied detector settings were produced and interpolated at the final level of the analysis, allowing the fitter to make arbitrary modifications to them.

In similar livetime as IC79-A and IC79-B, IC86-A selected 1487 neutrino events for analysis. While the best fit obtained was in agreement with the other results, the error in $\Delta m^2_{32}$ was reduced by about 20\% with respect to IC79-A, while maintaining a similar precision on $\sin^2 2\theta_{23}$. Figure~\ref{fig:icecube_yanez1} shows a comparison of data and best-fit simulation in projections in energy of the two-dimensional histogram used in the analysis. A comparison of the confidence regions in $\sin^2\theta_{23}$ and $\Delta m^2_{32}$ of the single year analyses of IceCube DeepCore, together with the result from ANTARES, is shown in Fig.\,\ref{fig:osc_compilation}.

\subsection{Precision measurements with IceCube DeepCore}
The latest result from IceCube DeepCore \cite{icecube_yanez} is an update to the IC86-A analysis introduced before, now with almost a thousand days of detector live time. The measurement demonstrates the potential for VLVNTs to become relevant experiments in the field of neutrino oscillations. 

\begin{figure*}[hbt]
 \includegraphics[width=0.9\textwidth]{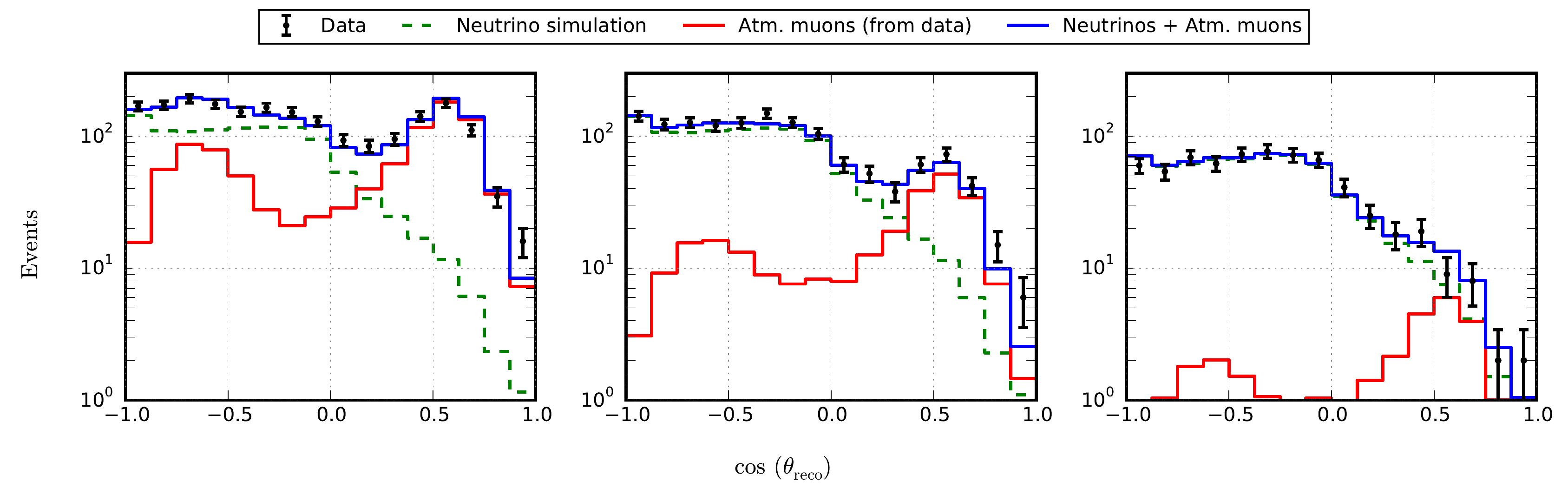}
 \caption{\label{fig:ic_bkg}Zenith angle distributions of neutrino simulation and atmospheric muons derived from data for three subsequent steps in the event selection with increasing veto cuts in IceCube's IC86-B analysis. A comparison is also made to a 10\,\% control sample of the data. Note that the region $\cos\theta_z > 0$ is not used in the final analysis of the data. Taken from \cite{icecube_yanez}.}
\end{figure*}

While the analysis strategy is still to focus on the selection on clear tracks, for which a core of direct photons can be identified, three large improvements are introduced, namely:
\begin{itemize}
\item An optimization of the event selection, which results in 40\% more events.
\item The cosmic muon background is derived from data (tagged muons), avoiding the need of computationally expensive model-dependent simulation.
\item An improved estimator of the energy deposited at the interaction point, which reduces the error on the total neutrino energy by more than 30\% at 20\,GeV.
\end{itemize}

A demonstration of how the data-derived background is used can be seen in Fig.\,\ref{fig:ic_bkg}, where the distribution of events as a function of reconstructed zenith angle at the final level and two earlier stages of the event selection is shown. At each step the cosmic muon background is more strongly suppressed. The contribution of atmospheric muons in the down-going region can be seen at all steps, including the final sample to be analyzed.

For their IC86-B result, the IceCube collaboration has expanded the list of possible sources of uncertainties considered. Non-DIS events are a non-negligible fraction of the sample at $E_\mathrm{reco} \leq 20$\,GeV, and additional cross sections uncertainties on these interactions (about 20\%) were also included. A possible shift of 5\% in the energy scale of hadronic showers was also taken into account.

In 950 days of live time, a total of 5174 events were observed, while 6830 were expected without oscillations. Note that the energy range of the search was reduced in comparison with IC86-A to $E_\mathrm{reco} = [7,56]\,$GeV. The data were analyzed in a full three-neutrino oscillation formalism, including the effects induced by matter as neutrinos cross the Earth. The parameters that best describe the data, assuming a normal mass ordering, are $\sin^2\theta_{23} = 0.53^{+0.09}_{-0.12}$ and $\Delta m^2_{32}=2.72^{+0.19}_{-0.20}\times 10^{-3}$\,eV$^2$. No significant preference was found for either the normal or inverted mass orderings. Purely statistical uncertainties are ${}^{+0.06}_{-0.08}$ for $\sin^2\theta_{23}$, and ${}^{+0.14}_{-0.15}\times 10^{-3}$\,eV$^2$ for $\Delta m^2_{32}$, from which it is deduced that statistical and systematic uncertainties have an almost equal impact on the result.

\begin{figure}[hbt]
 \includegraphics[width=1.\columnwidth]{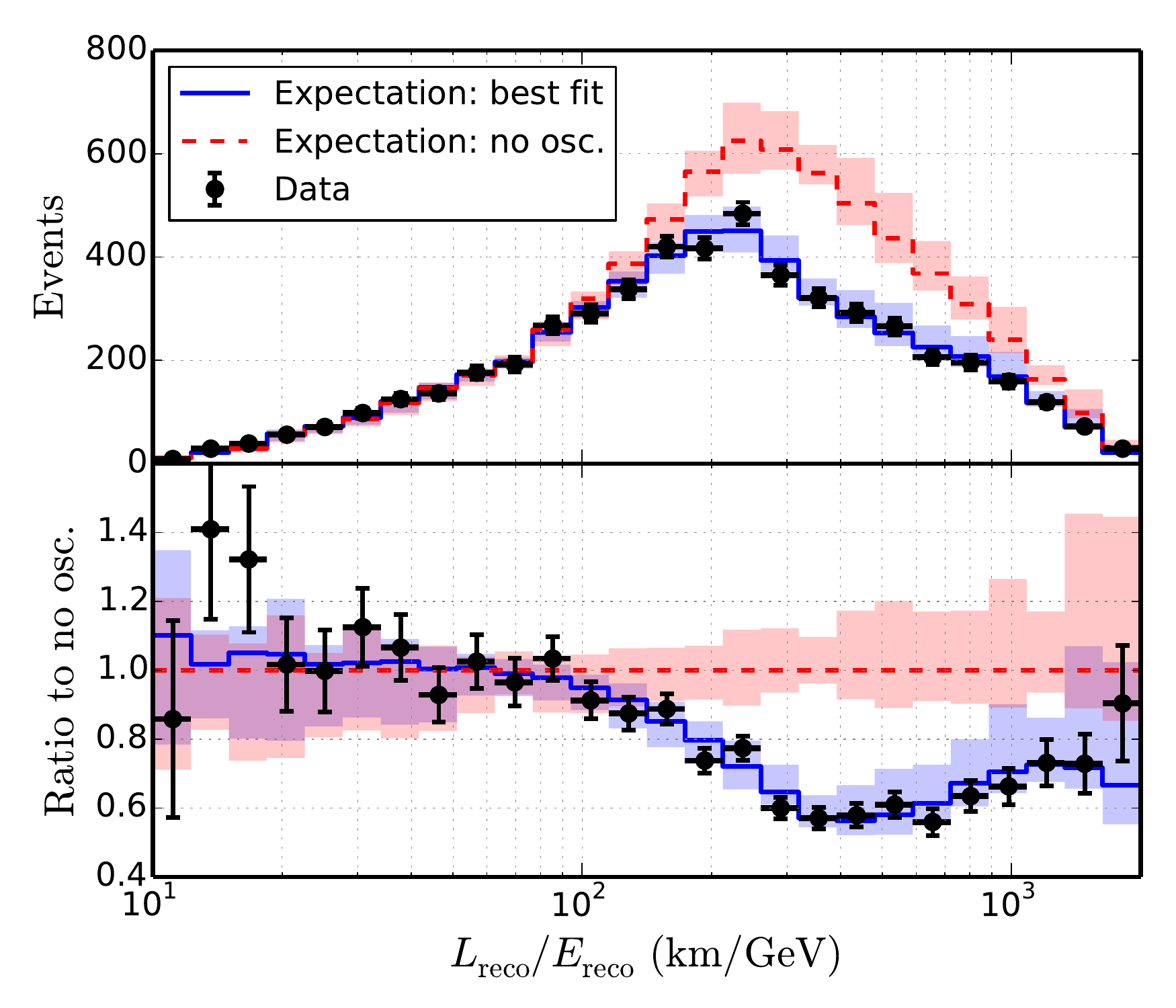}
 \caption{\label{fig:ic_LE}Distribution of events as a function of reconstructed $L/E$ of IceCube's IC86-B analysis. Data are compared to the best fit and expectation with no oscillations (top) and the ratio of data and best fit to the expectation without oscillations is also shown (bottom). Bands indicate estimated systematic uncertainties. From \cite{icecube_yanez}.}
\end{figure}

\begin{figure}[hbt]
 \includegraphics[width=1.\columnwidth]{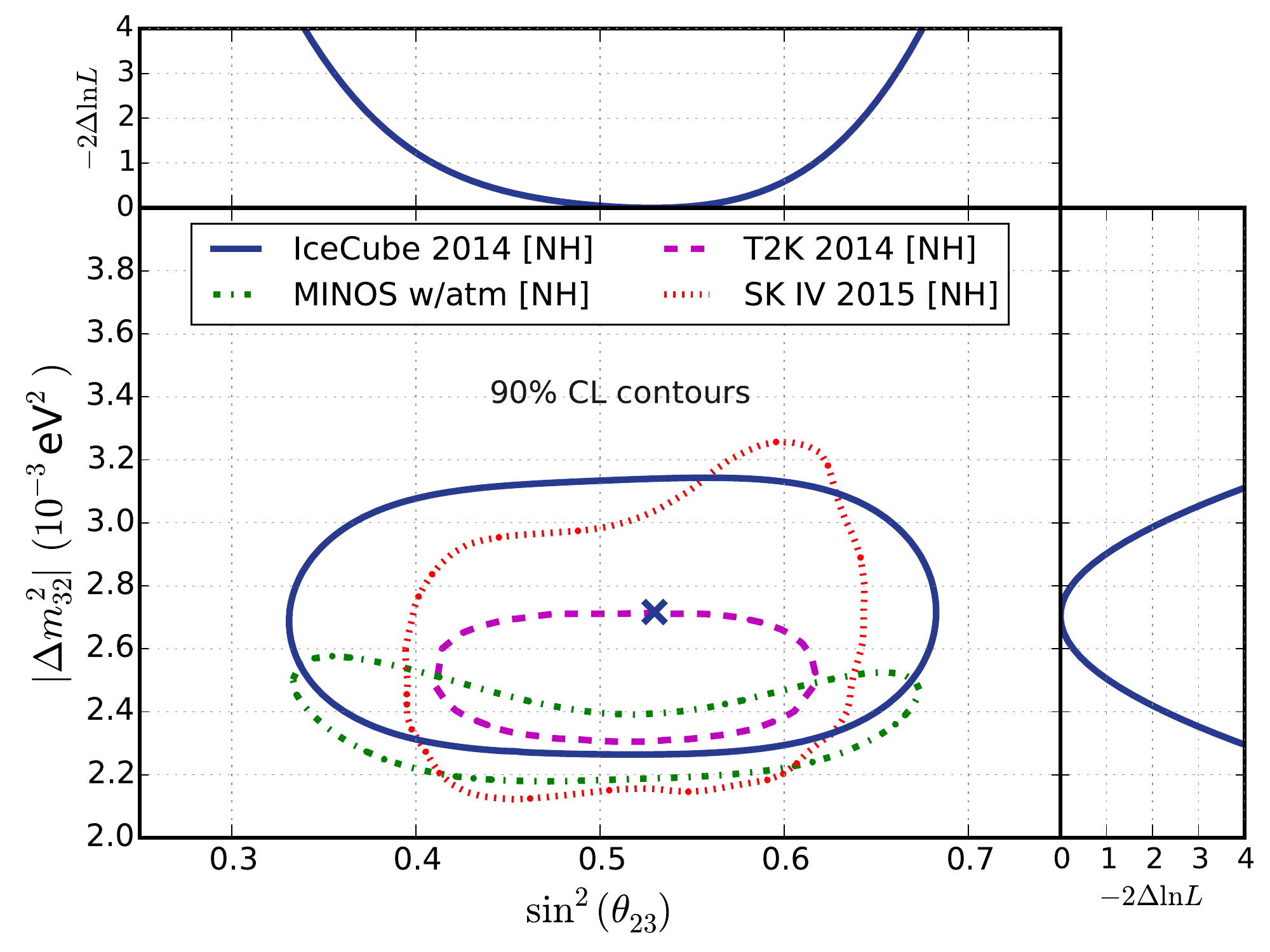}
 \caption{\label{fig:ic_contours}90\,\% confidence contours of the latest result from IceCube (IC86-B) in the $\sin^2\theta_{23}-\Delta m^2_{32}$ plane in comparison with the ones of the most sensitive experiments \cite{t2k_disappearance, minos1, SK2014}. The log-likelihood profiles for individual oscillation parameters are also shown (right and top). A normal mass ordering is assumed. Updated from \cite{icecube_yanez}.}
\end{figure}

Data and simulation are in good agreement, with a $\chi^2$/NDF$=54.9/56$ for the energy-zenith angle histogram used in the fit. Figure\,\ref{fig:ic_LE} compares the $L_\mathrm{reco}/E_\mathrm{reco}$ distributions of data and best fit simulation, where the agreement can be observed\footnote{Note that the analysis is not done on this variable, but in a two-dimensional energy-zenith angle histogram instead.}. The 90\,\% confidence contours on the atmospheric oscillation parameters obtained are shown in Fig.\,\ref{fig:ic_contours}, together with the results from the other experiments leading the field. 

The results from VLVNTs will be further improved by adding statistics to the analyzed data sample and refining the reconstruction methods. However, the most decisive improvements will come with the construction of the next-generation VLVNTs presented in the next section.

\section{Neutrino oscillations with the next generation of VLVNTs}
\label{sec:future}

After the measurements from ANTARES and IceCube/DeepCore in the atmospheric sector, the next goal of VLVNTs is to further decrease the energy threshold below the 15\,GeV domain in order to improve the sensitivity to the PMNS matrix elements and determine the NMO. Measuring the neutrino mass ordering is the main objective of the forthcoming ORCA (Oscillation Research with Cosmics in the Abyss) \cite{ORCA_ICRC} and PINGU (Precision IceCube Next Generation Upgrade) \cite{pingu_koskinen, pingu} detectors as part of the KM3NeT \cite{km3net} and IceCube Gen-2 \cite{icecube_gen2} infrastructures, respectively.

\subsection{Design of future detectors}

Both ORCA and PINGU will be more densely equipped than the currently operating detectors and should reach several megatons in instrumented volume. Their concepts are similar in many ways, with the most significant differences coming from the detection medium, the proposed detector layout and the (default) optical module design.

\subsubsection{Hardware and detector geometry}

\begin{figure}[bt!]
\centering
 \includegraphics[width=0.99\columnwidth]{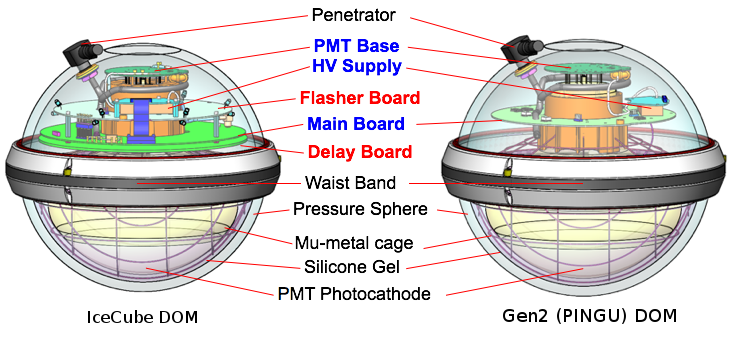}
 \caption{Comparison between the currently operating IceCube DOM and the updated PINGU/Gen2 DOM.}
 \label{fig:PINGU_DOM}
\end{figure}

\begin{figure}[bt!]
\centering
 \includegraphics[width=0.9\columnwidth]{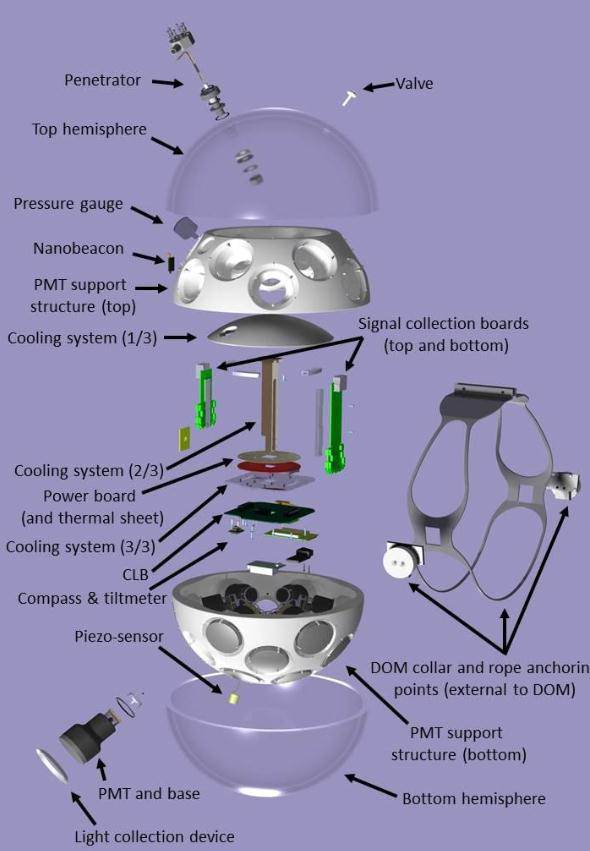}
 \caption{An exploded view of the multi-PMT optical module of KM3NeT/ORCA.}
 \label{fig:ORCA_OM}
\end{figure}

The PINGU optical module will most likely be a simplified and modernized version of that of IceCube, which has demonstrated its stability and reliability over almost ten years of operation. The PINGU DOM design removes components that are no longer required, such as the local coincidence logic and the multiple amplification modes, while providing a larger dynamic range than the original IceCube DOM and improved time resolution of 2\,ns \cite{pingu}. A schematic view of the IceCube and PINGU (Gen2) DOMs is shown in Fig.~\ref{fig:PINGU_DOM}. By maintaining the basic IceCube design, the PINGU DOM minimizes risk and cost. The ORCA optical module will follow the KM3NeT design \cite{km3net} with each DOM housing 31 small (3") PMTs arranged in a 17" glass sphere together with the associated electronics, as can be see from Fig.~\ref{fig:ORCA_OM}. This design offers the possibility of creating coincidences within the OM to suppress the large $^{40}$K decay background as well as the thermal noise of the PMTs. The orientation of the PMTs within the OM is also used in the reconstruction of events, although not yet at its full potential. A single sphere houses three to four times the photo cathode area of an ANTARES OM with an almost uniform angular coverage, improving the cost effectiveness by a factor four. Several prototypes of such a multi-PMT OM have been successfully tested in-situ \cite{km3net_om}.

The final layouts of ORCA and PINGU are still under optimization\footnote{Preliminary results tend to indicate that the best vertical spacing between OM is around 10~m for ORCA, while similar studies in the PINGU case favor a vertical spacing of about 3~m, close to the adopted benchmark.}. The current benchmark geometries used for establishing the detector performances consist of 40 (115) strings with a horizontal spacing of $\sim$20~m for PINGU (ORCA). The vertical spacing is set to 6~m for ORCA and 3~m for PINGU. While a PINGU string will hold up to 96 DOMs, there are 18 DOMs in a default ORCA string. The maximum number of DOMs that a PINGU string can hold is given by the mechanical constraints of the down-hole cable and the appearance of shadowing effects, while for ORCA the constraint comes from the launcher vehicle (a large spherical frame in which the DOMs slot into dedicated cavities) used for string deployments. The separation between the sensors of both detectors is smaller than the absorption and scattering lengths of their respective media, making the optical properties of ice and salt water less relevant than for ANTARES and IceCube/DeepCore.

\begin{figure}[bt!]
 \centering
 \includegraphics[width=0.4\textwidth]{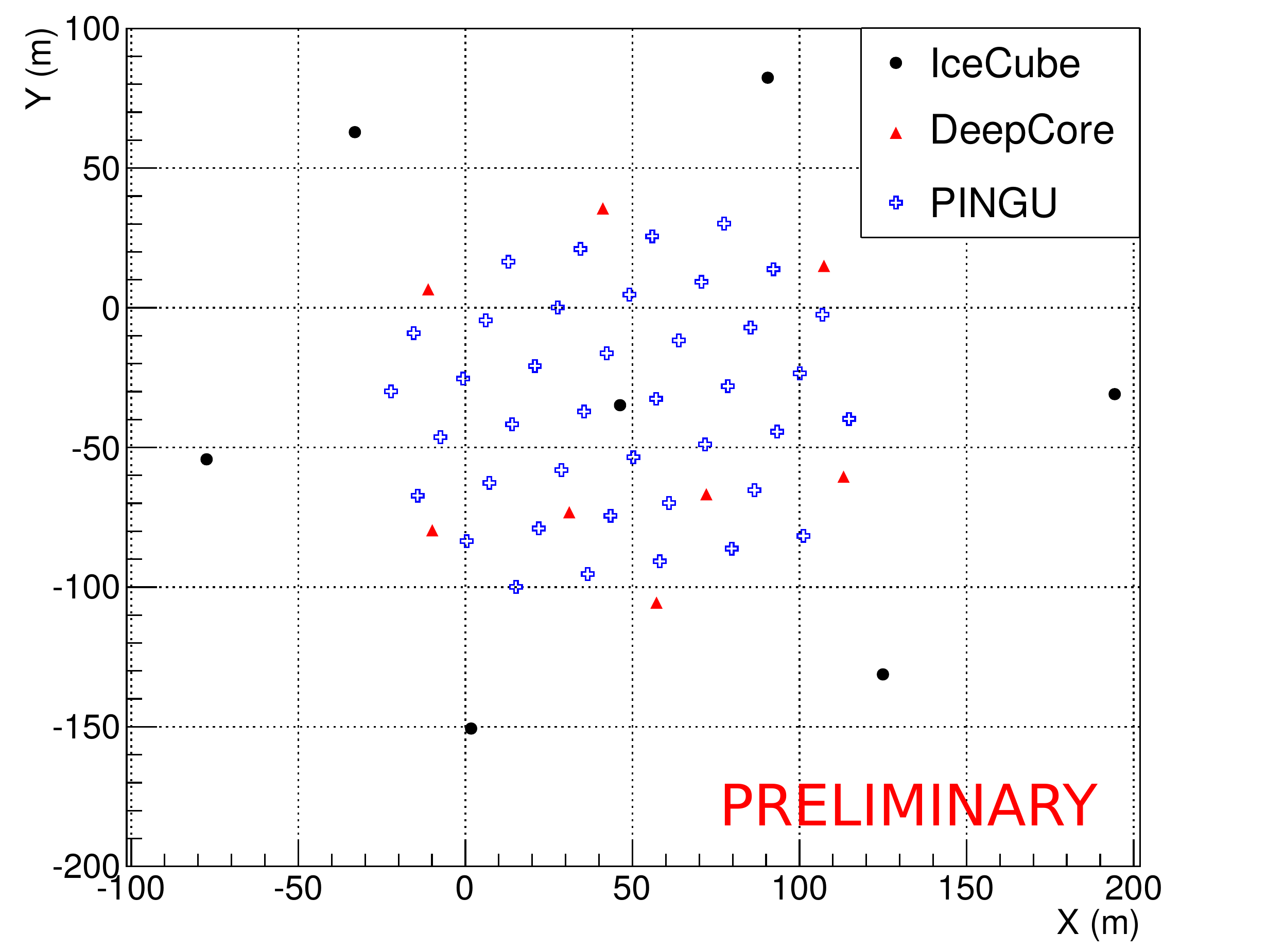}\\
 \centering\includegraphics[width=0.4\textwidth]{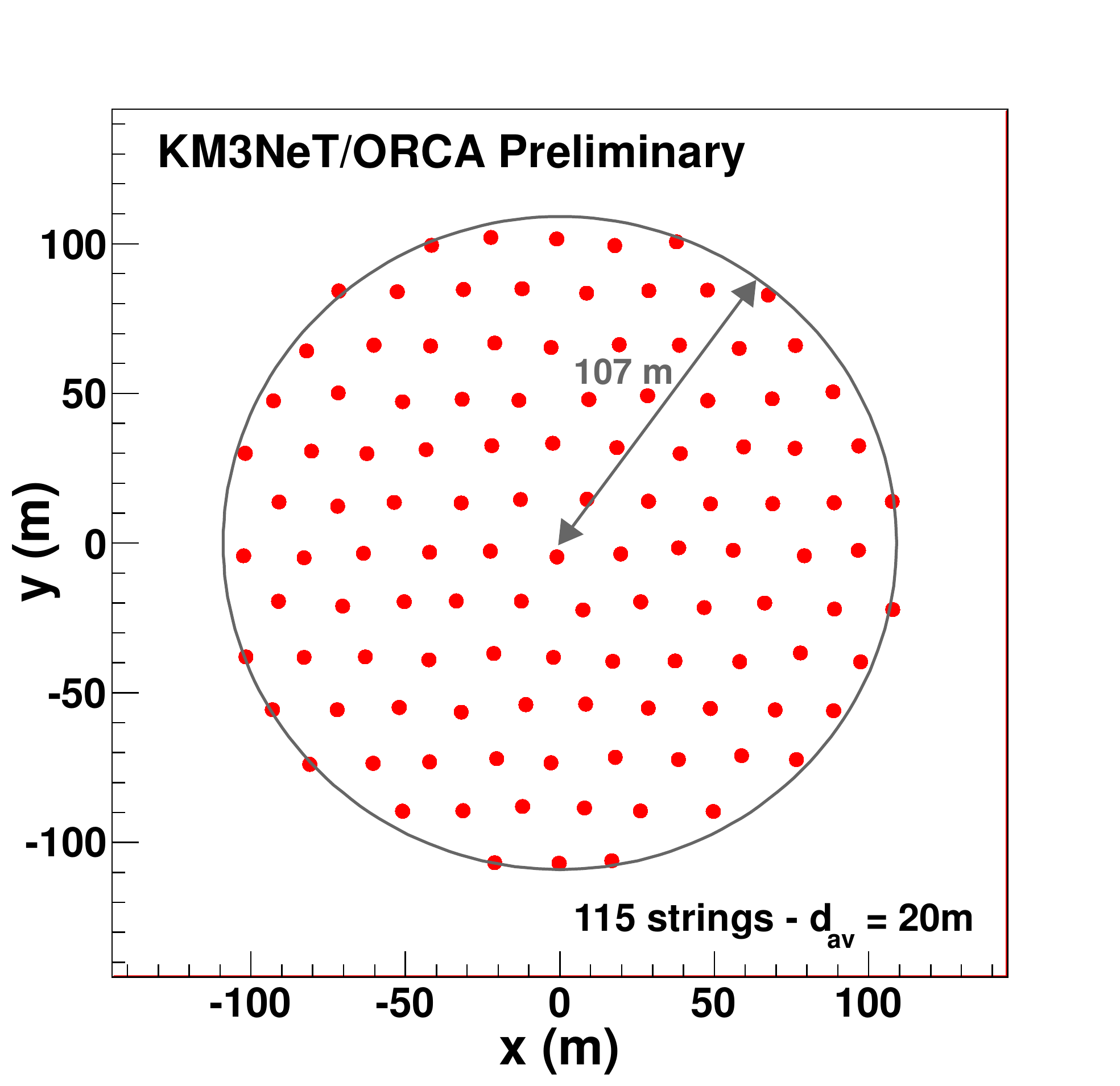}
 \caption{Top: an envisaged 40-string PINGU layout (blue strings). The black circles refer to the standard IceCube strings, the red triangles to the DeepCore strings. Bottom: ORCA benchmark detector footprint.}
 \label{fig:footprints}
\end{figure}

The footprints of the ORCA and PINGU detectors are shown in Figure~\ref{fig:footprints}. The instrumented mass of both detectors is of order 3.5 to 4~Mt, and their effective masses reach the same value for neutrinos of energy above 10 GeV. While the PINGU extension is foreseen to be embedded inside the current IceCube/DeepCore detector (which will be used for background vetoing), the ORCA detector will be located around 10~km west from the ANTARES site, at a depth of 2475m.

\subsubsection{Costs and timescale}
PINGU estimates a cost of 48M\$ for hardware and 23M\$ for logistics \cite{ty_dpf}. The estimated cost of ORCA is 40M\euro. Funding request processes are currently driving the possible time line of the projects. 

PINGU will be built as part of the IceCube Gen-2 project. From a technical point of view the installation of the detector at South Pole could start by the end of 2020 \cite{hanson_vlvnt}. Based on the experience gained with the IceCube, the deployment is expected to take only three years. The first construction phase of ORCA, a demonstrator array of 6-7 strings (already funded), started in late 2014 with the deployment of the main electro-optical cable, followed by the deployment of a junction box in April 2015. The demonstrator is expected to be deployed by the end of 2016, and will be used to carry out studies of detector-related systematic effects and event reconstructions. In an optimistic case, the deployment of the full detector case could happen by 2020. Both PINGU and ORCA plan to take data during their construction phase.

\subsection{Projected performance}

The determination of the NMO, the main physics goal of these projects, relies on a detailed analysis of deviations of the order of $\sim$10\% and $\sim30$\% in the rates of detected atmospheric muon and electron neutrinos (see Figs.~\ref{fig:osc_prob1}, \ref{fig:osc_prob2}, \ref{fig:osc_prob3}) as a function of energy and arrival zenith angle. Therefore, the key parameters that characterize the potential of a detector are its effective mass, the energy and zenith angle resolutions achievable, and its particle (mis-)identification capabilities. In the following discussion the latest, preliminary, studies from ORCA~\cite{ORCA_ICRC2, ORCA_ICRC} and PINGU~\cite{pingu,PINGU_ICRC} are presented.

These studies are based on full Monte Carlo simulations adapted from IceCube and ANTARES. All ORCA results account for an optical background induced by $^{40}$K decays of 5-10 kHz per PMT and a time-correlated hit rate of 500 Hz per OM (two coincident hits in different PMTs inside the same OM). Since PINGU DOMs will follow closely the design used for IceCube, the typical in situ behaviour of the IceCube/DeepCore DOMs, with a noise rate of 650\,Hz, is used in the simulations.

The published results of ANTARES and IceCube have so far focused on $\nu_\mu$ disappearance, and therefore only selected events where a muon was observed. The sensitivity to the NMO, on the other hand, also comes from oscillations that involve $\nu_e$. It is therefore useful to detect all neutrino flavors, placing them in two categories depending on their topology: tracks and cascades (see Sec.~\ref{sec:pid}).

\subsubsection{Reconstruction of tracks and cascades}

Track-like events are those where a muon is observed coming out of the interaction vertex. Track-like topologies are CC $\nu_\mu$ interactions as well as the $\nu_\tau$ CC interactions when the decay of the tau lepton produces a muon. The cascade-like topologies are CC $\nu_e$ interactions, CC $\nu_\tau$ interactions without a muon in the final state, and NC interactions from all flavors. Independent studies indicate that after accounting for reasonable detector resolution effects, the cascade channel provides more sensitivity to the effects of the NMO. Note, however, that the two channels are complementary as track-like events can provide better precision in $\sin^2 \theta_{23}$. It is consequently important to be able to distinguish the two topologies with high efficiency and purity.

The event reconstruction in PINGU is a simultaneous global likelihood fit of the interaction vertex position and time, the zenithal and azimuthal angles, the energy of the cascade at the vertex, and the length of the daughter muon track. The event hypothesis assumes that tracks and cascades are collinear. The likelihood is calculated using the time of arrival of single photons and the expected noise in the time windows analyzed. The expectations for minimum ionizing muon tracks and electromagnetic cascades needed for the likelihood are stored in tables, obtained from direct simulation of particle and photon propagation, as it is already done for IceCube \cite{icecube_energy}. An event is reconstructed by comparing photon expectation for a given event hypothesis to the photons observed. All the DOMs in PINGU, as well as those in IceCube/DeepCore, are used in the reconstruction \cite{pingu}.

\begin{figure}[bt!]
\centering
 \includegraphics[width=0.49\textwidth]{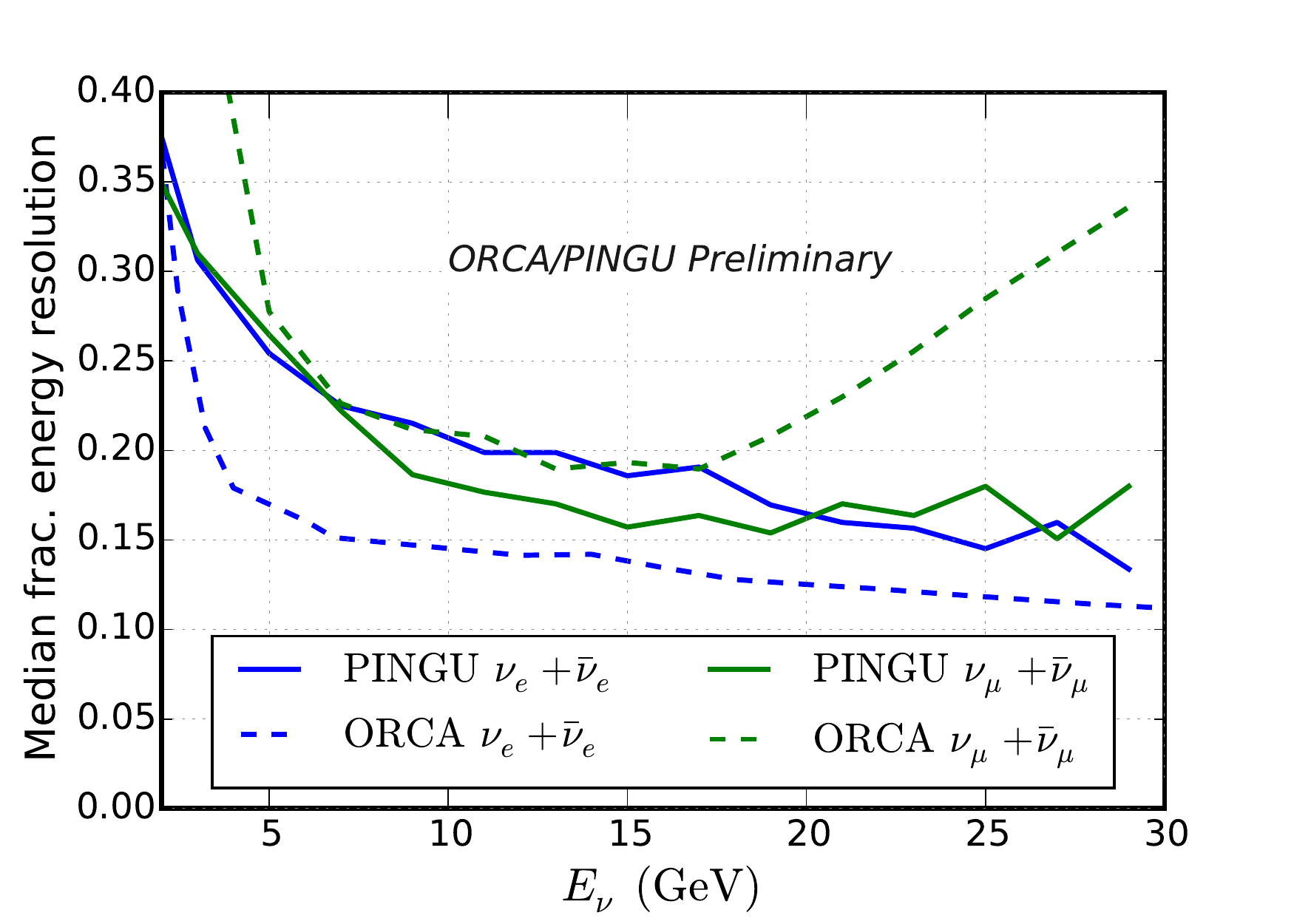}
 \caption{Expected median fractional energy resolution for electron and muon neutrinos in PINGU (solid) and ORCA (dashed). Reproduced from \cite{pingu, ORCA_ICRC}.}
 \label{fig:ShowerReso1}
\end{figure}

\begin{figure}[bt!]
\centering
 \includegraphics[width=0.49\textwidth]{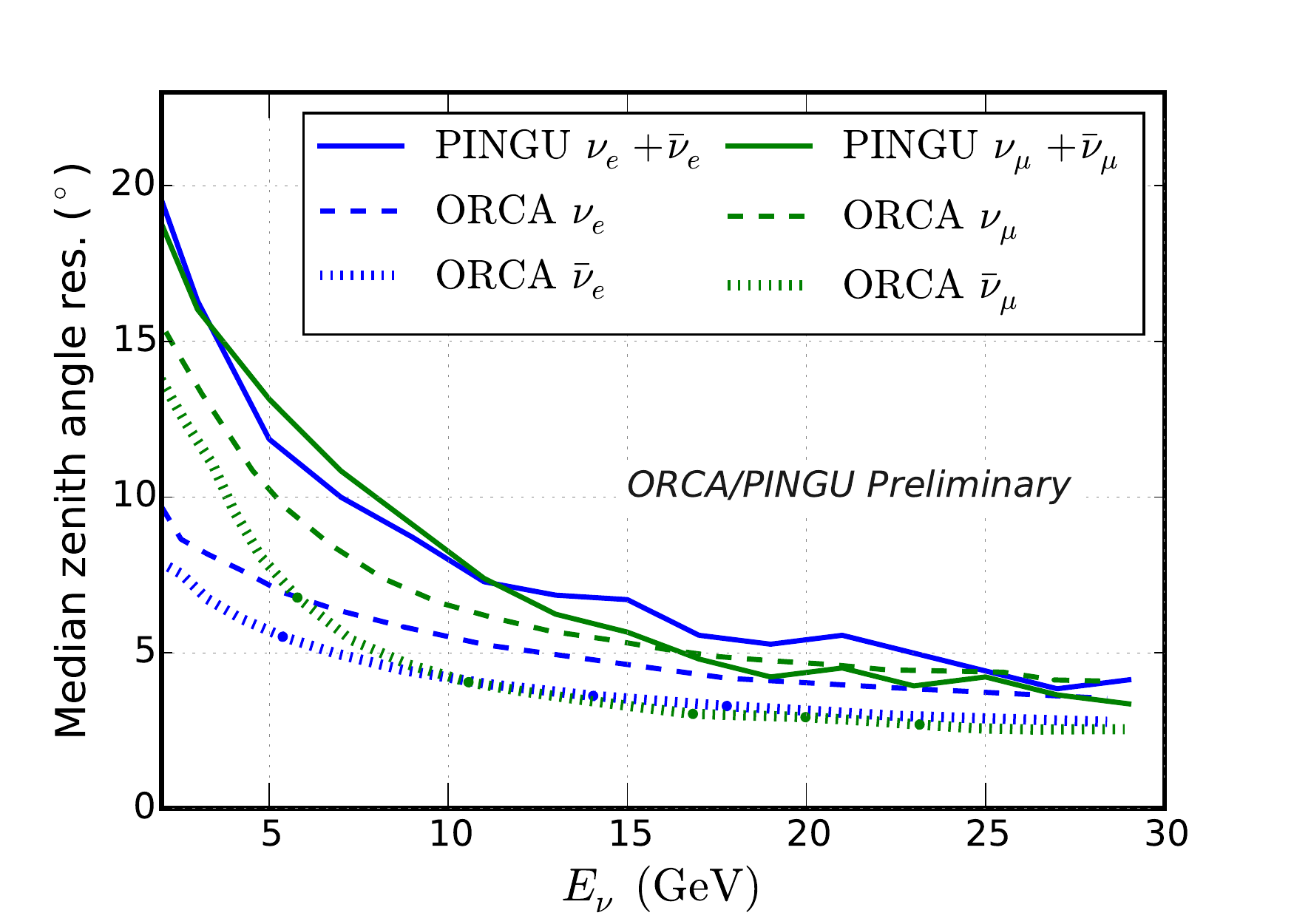}
 \caption{Expected median zenith angle resolution for electron and muon neutrinos in PINGU (solid) and ORCA (dashed). For ORCA individual resolutions for neutrinos and antineutrinos are shown, while a mixture of both is given for PINGU. Resolutions are better for anti-neutrinos than for neutrinos due to the smaller average inelasticity, leading to a smaller intrinsic scattering angle between the neutrino and the leading lepton. Values taken from \cite{pingu, ORCA_ICRC}.}
 \label{fig:ShowerReso2}
\end{figure}

Fitting eight parameters at once while simultaneously looking up expectations from tables makes the reconstruction CPU intensive, but in return it provides robust results and similar resolutions for track-like and cascade-like topologies. While it would be possible to use the information provided by this reconstruction to obtain an estimate of the inelasticity of the event, this has not been explored so far. Energy and zenith angle resolutions for different interactions are shown in Figs.~\ref{fig:ShowerReso1} and~\ref{fig:ShowerReso2}, together with those obtained by ORCA with the methods explained hereunder.

ORCA uses two distinct algorithms for tracks and cascades. The track reconstruction is directly adapted from the main reconstruction of ANTARES \cite{antares_reco} and focuses on the muon direction using the combined information of the PMT spatial positions and the Cherenkov photon arrival times. The neutrino energy estimation is mainly given by the reconstructed muon track length, which is complemented by the number of hits used in the track reconstruction algorithm. Muon tracks produced in neutrino interactions at $E_\nu\geq$15\,GeV are not always fully contained, which turns the estimate into a lower limit above these energies, as shown in Fig.~\ref{fig:ShowerReso1}. The time residuals under a spherical emission profile (shower-like) or according to a Cherenkov cone (track-like) are used to obtain sensitivity to the inelasticity in the track channel.

\begin{figure}[bt!]
\centering
 \includegraphics[width=0.45\textwidth]{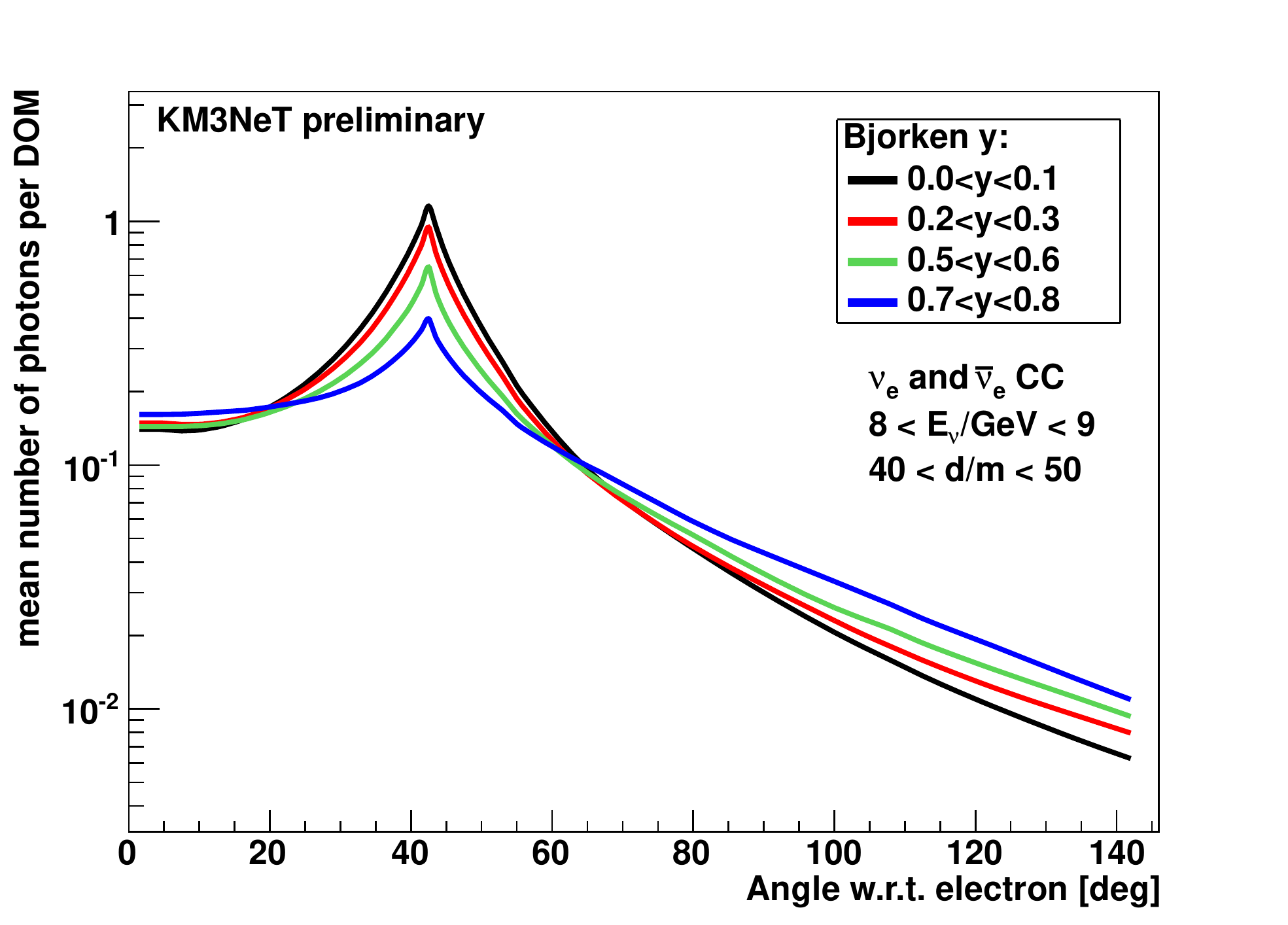}
 \caption{ \label{Fig:CherenkovShower} Number of expected photons as a function of the emission angle between the shower direction and the direction from the vertex to the DOM for different intervals of interaction inelasticity $y$.}
\end{figure}

The cascade reconstruction in ORCA takes advantage of the long scattering length in sea water, which preserves the structure of the Cherenkov light cone, and tries to identify the leading lepton in the cascade. An example of the distribution of the expected number of photons as a function of emission angle for different inelasticity intervals is shown in Fig.~\ref{Fig:CherenkovShower}. A peak is always visible at the Cherenkov angle (42$^\circ$), whose height with respect to the off-peak region depends on $y$. Cascades are reconstructed in two separate steps using maximum likelihood fits. First the interaction vertex is obtained with a resolution of about 0.5-1~m by an algorithm based on hit time residuals. It is then followed by a fit of the direction, energy and inelasticity of the event. The performances of the cascade reconstruction are summarized in the Figures~\ref{fig:ShowerReso1} and~\ref{fig:ShowerReso2}. 

In ORCA the inelasticity of about 60\% of the tracks with true $y\leq0.25$ or $y\geq0.75$ is reconstructed correctly; the accuracy of the inelasticity estimator of cascades is slightly worse. The inelasticity could be used for potential statistical separation between neutrinos and anti-neutrinos, which can be exploited for the mass ordering measurement \cite{nmh_y}. It can also be tested to separate charged current interactions from neutral current interactions. While both PINGU and ORCA are studying this possibility, inelasticity estimates are not yet part of the current analyses that are discussed in the following sections.

\subsubsection{Particle identification and background rejection}
\label{sec:pid}
VLVNTs measuring atmospheric neutrinos should be able to identify and reject atmospheric muons, the largest source of background, and differentiate between events with track-like and cascade-like topologies. PINGU plans to tag atmospheric muons following the strategy developed in DeepCore, i.e. using the outer detector strings to identify particles that enter the fiducial volume, and restricting the analysis to starting and up-going events (see \cite{icecube_yanez} and Fig.~\ref{fig:ic_bkg}). The cosmic muon background is expected to be on the level of a few percent, similar to DeepCore. Event reconstruction and selection in PINGU do not rely on direct hits, the single largest impact on signal efficiency in the latest DeepCore results. Signal efficiency in PINGU, therefore, is expected to be minimally affected by background rejection and reconstruction methods, and largely defined by the number of photons observed from an interaction.

The ORCA detector does not rely on an outer detector to tag muons. Current analyses reduce the impact of these muons by selecting only up-going events and rejecting the mis-reconstructed ones using variables such as their reconstruction quality and the position of their reconstructed interaction vertex. The topology of neutrino interactions, track-like or cascade-like, is identified using the distribution of hit time residuals, distances between reconstructed vertices at various reconstruction steps, the quality of the reconstructions, and topological variables, among others. A single multivariate method which incorporates the parameters listed above is applied to the data, and classifies events as tracks, showers or atmospheric muons. The procedure achieves $\sim$1\% muon contamination in the final sample without a severe signal loss.

\begin{figure}[bt!]
\centering
 \includegraphics[width=0.49\textwidth, trim=0 0 30 0]{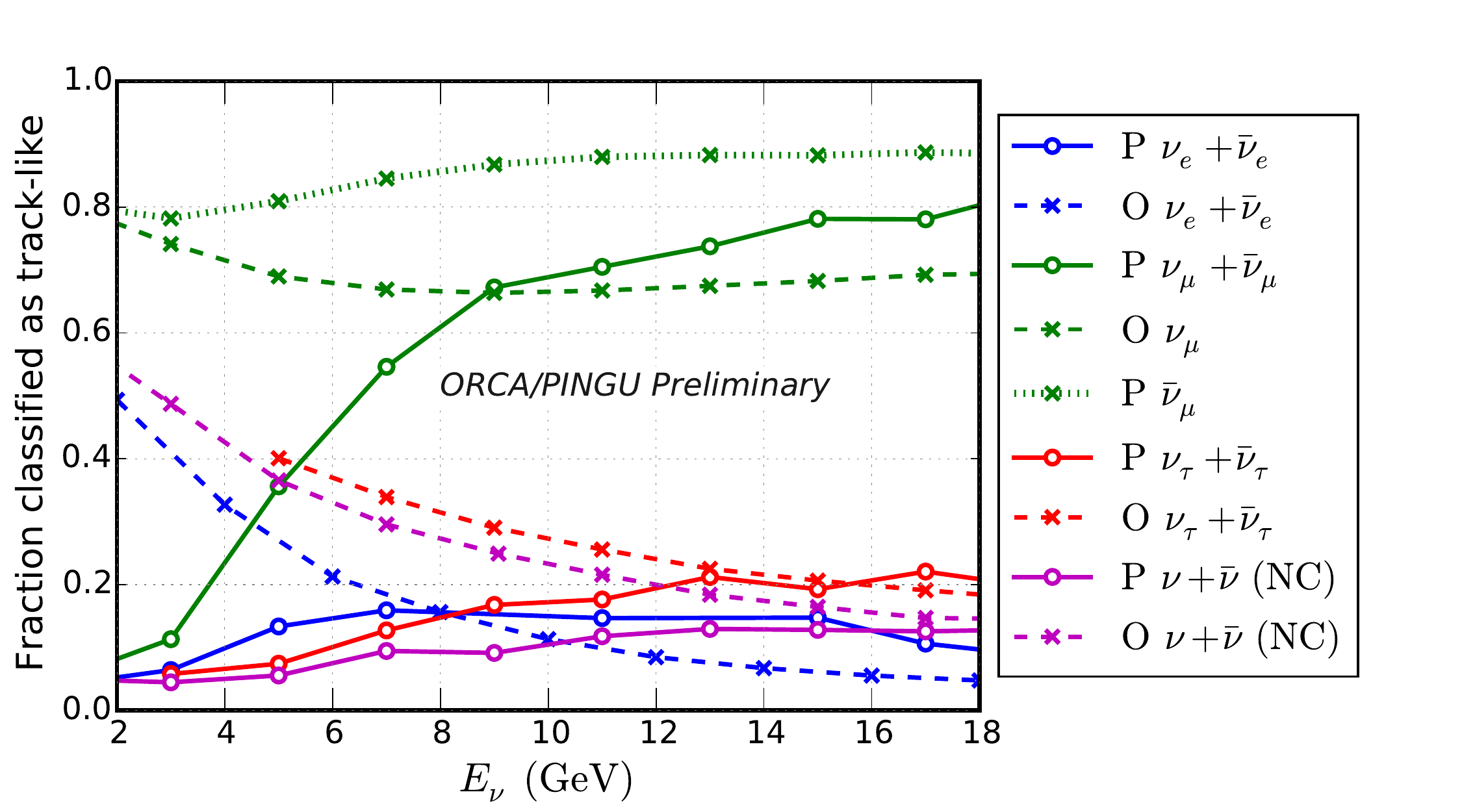}
 \caption{Fraction of events selected as tracks for different categories of simulated events for ORCA (dashed lines, labeled as O) and PINGU (solid lines, labeled as P). As expected, $\bar{\nu}_\mu$ performs better because of their average lower inelasticity. From \cite{jp_vlvnt}. }
 \label{fig:PID}
\end{figure}

The approach followed in PINGU to separate tracks from cascades also uses a multivariate method with variables describing the reconstruction quality of the event under the track vs cascade hypothesis, as well as the reconstructed muon track length as input. Figure~\ref{fig:PID} shows a comparison of the performance of these classification methods for neutrino interactions around the energies relevant for mass ordering measurements for PINGU and ORCA. Current methods differ at low energies, with PINGU showing a bias towards classifying low energy tracks as cascades and ORCA exhibiting the opposite behavior. Above 10\,GeV both classification schemes result in a similar outcome. The results suggest that the behavior of the particle identification algorithms at low energy can be tuned for optimizing sensitivity to the NMO measurement. In both cases, the final performances are subject to further optimization.

\subsection{Physics potential and systematics}

The preliminary performances described above are used by the PINGU and ORCA collaborations as inputs to estimate the confidence level with which the projected experiments will be able to reject a given NMO. This is done by drawing several thousands of pseudo-experiments generated under each mass ordering hypothesis, as outlined in \cite{franco_nmh}. The analysis is conducted by comparing the two-dimensional histograms of pseudo-data and simulation as a function of the reconstructed energy and zenith. The pseudo-data sets are generated varying different parameters, such as the values of the mixing angles, in order to study the impact of degeneracies in the measurement.

\begin{table}[t]
\centering
   \caption{\label{tab:sys} List of the uncertainties studied by ORCA and PINGU which have the largest impact on their respective NMO analyses (more systematic uncertainties have been studied, see text). Sources of uncertainty are additional parameters in the fit. Studies are performed for a set of true oscillation parameters. The best known values for all other parameters are injected for creating the data templates. PINGU uses priors to penalize deviations while fitting these parameters. ORCA does not use priors, and instead reports the standard deviation of the fit results.}
\begin{Tabular}[1.3]{l|c|c}
  \hline
  Uncertainties &  ORCA     & PINGU \\
   & $\sigma$(fit yield)      & $\sigma$(prior) \\
   \hline \hline
   $\theta_{23},\,\Delta m^2_\mathrm{31}$ &   \multicolumn{2}{c}{Unconstrained} \\\hline
   $\theta_{13}$ & Integrated $\pm 1^\circ$& 0.2$^\circ$ \\\hline
   $\theta_{12},\,\Delta m^2_\mathrm{21}$ & \multicolumn{2}{c}{Fixed} \\\hline
   $\delta_\mathrm{CP} $& \multicolumn{2}{c}{Fixed at zero\footnote{Both projects have studied how $\delta_\mathrm{CP}$ impacts their sensitivity but the results are not yet reflected in the projections given in this review.}}\\ \hline\hline
   Overall rate factor &2.0\%& Unconstrained \\ \hline
    $E^{-\gamma}$ (slope, spectral index) & 0.5\% & $\pm$0.05 \\ \hline
    Energy scale  & Not used &  $\pm$10\% \\ \hline
    $\nu$ / $\bar{\nu}$ ratio & 4.0\% & $\pm$10\%  \\\hline
   $\mu$ / e  flavor ratio & 1.2\%& $\pm$3\%  \\\hline
   NC cross section scaling & 11.0\%& GENIE model\\\hline
  \hline
\end{Tabular}
\end{table}

A full log-likelihood ratio (LLR) method is used by both collaborations to report their expected sensitivity. In this method each pseudo-experiment is analyzed by performing  a log-likelihood fit with the oscillation parameters as free parameters (mostly $\theta_{23}$, $\Delta m ^2 _{32}$ and $\theta_{13}$) and assuming both hierarchies in turn. Sources of systematic uncertainty are incorporated as additional parameters in the fit (see Table~\ref{tab:sys}).

As such methods can be quite CPU expensive, in particular when studying various sources of systematics, the PINGU collaboration also implemented a simplified $\Delta \chi^2$-based approach. This method is a parametric analysis based on the Fisher information matrix, which relies on the partial derivatives of the event counts in each bin with respect to all parameters under study. Inverting the Fisher matrix yields the full covariance matrix between the parameters. The covariance matrix of the mixing angle $\theta_{23}$ is calculated at several values to overcome the limitations of the method. The results obtained with the Fisher matrix are in agreement with the LLR method, and are also used to report the projected sensitivity of PINGU.

\begin{figure}[bt!]
\centering
 \includegraphics[width=0.49\textwidth]{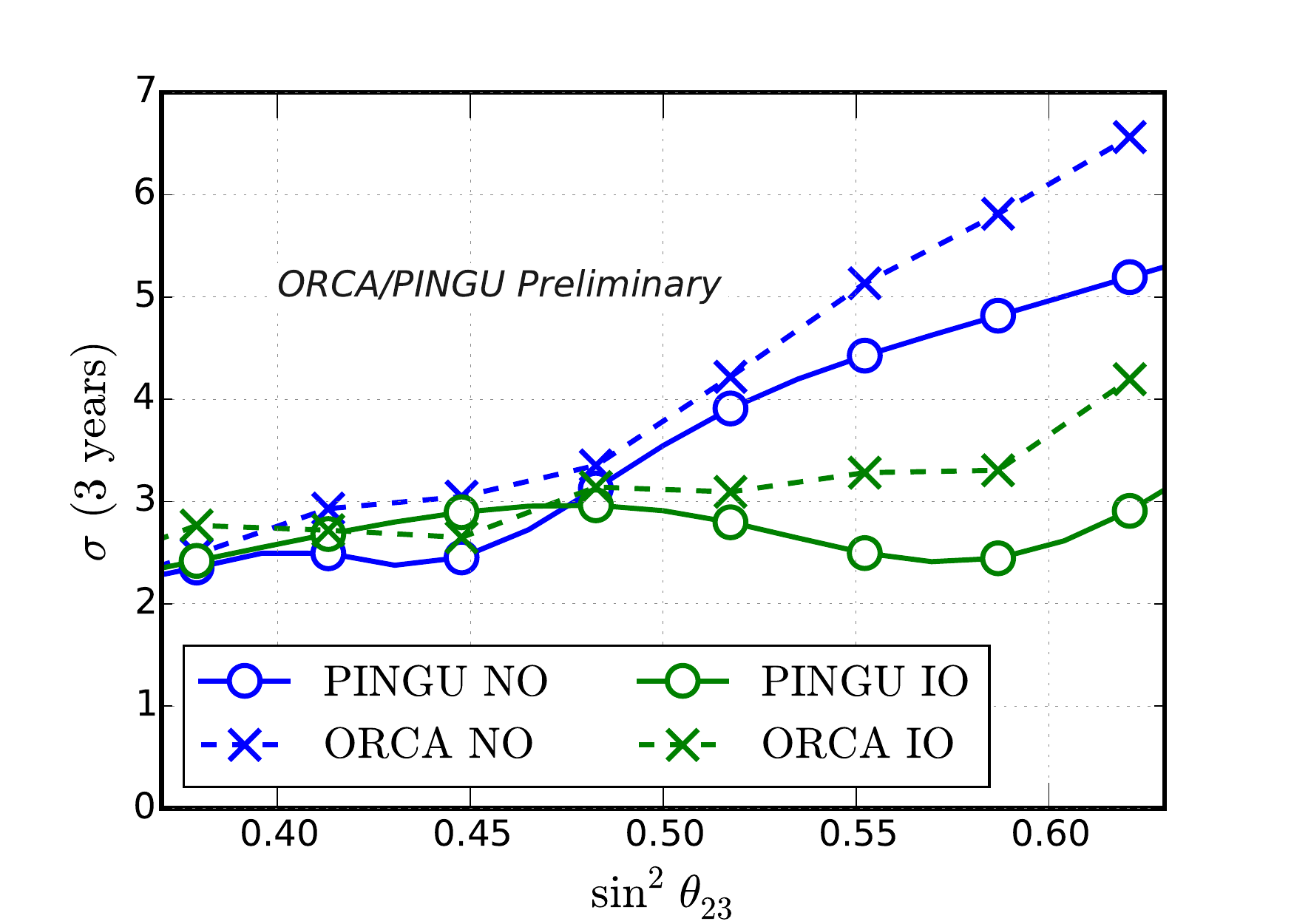}
 \includegraphics[width=0.49\textwidth]{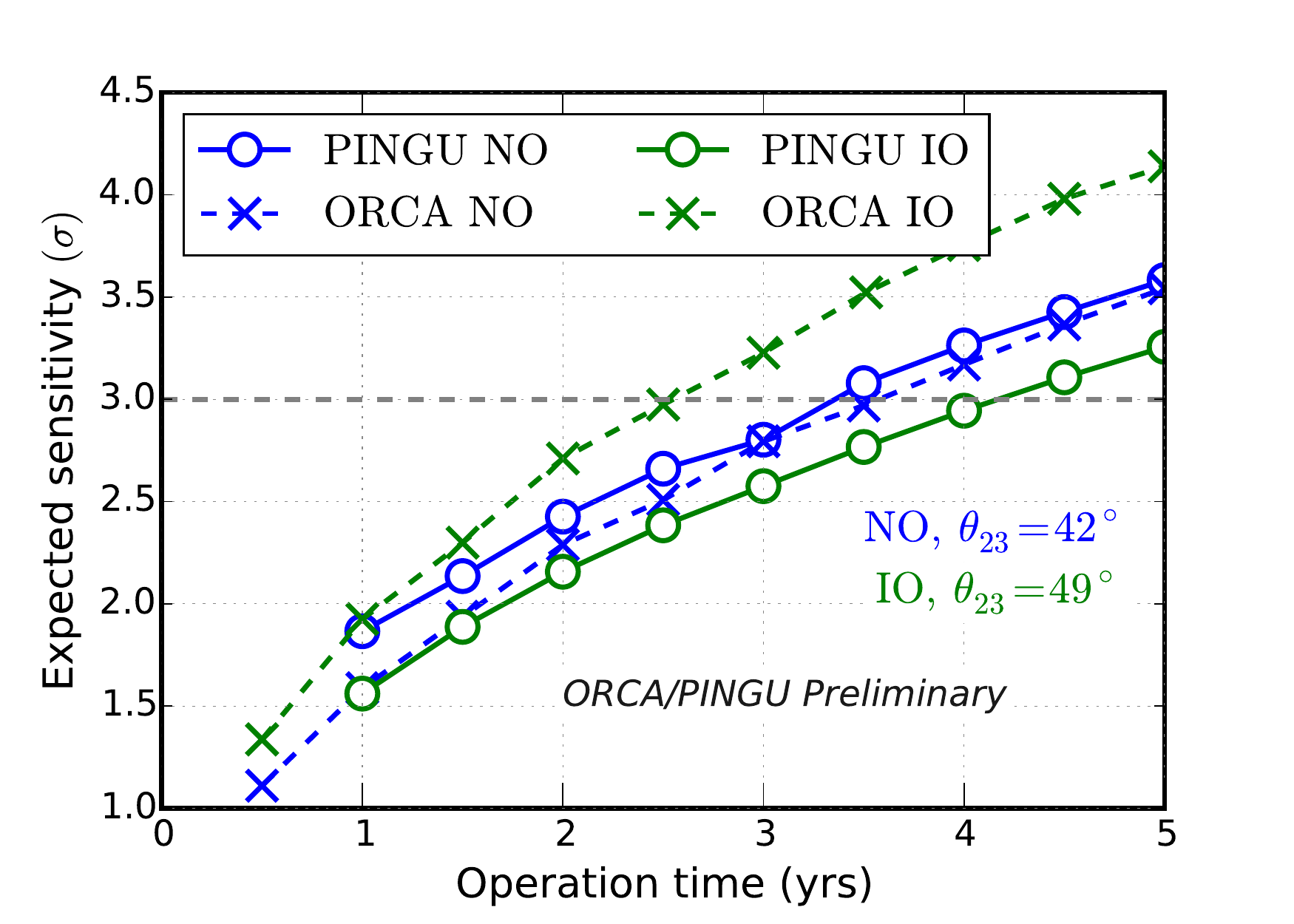}
 \caption{  \label{fig:sensNMO} Top: Significances of ORCA and PINGU for rejecting a given hypothesis for the neutrino mass ordering plotted as a function of $\theta_{23}$, after 3 years of data taking. Bottom: Median significances as a function of time for the benchmark detectors described in the text. The oscillation parameters injected are close to those found in \cite{Gonzalez-Garcia:2014bfa} ($\theta_{23}=42^{\circ}$ for a NMO, $\theta_{23} = 49^\circ$ for an IMO). From \cite{ORCA_ICRC, PINGU_ICRC}.}

\end{figure}

The parameters of the fits performed by ORCA and PINGU, presented in Table~\ref{tab:sys}, are the oscillation parameters of interest plus a set of parameters related to uncertainties on the detection process, neutrino fluxes, cross sections and the remaining oscillation parameters. The oscillation parameters, in particular $\theta_{23}$, have the largest impact on the achievable precision. The overall normalization has the second largest impact on the precision. This absorbs uncertainties on the efficiency of the detector, the absolute atmospheric neutrino flux and interaction cross sections. PINGU has recently studied uncertainties on the neutrino flux by using a more refined description, which involves a set of 18 parameters \cite{Barr:2006it}. The impact found was a reduction of the three year sensitivity by 0.2$\sigma$ \cite{joakim_vlvnt} (not yet included in Fig.~\ref{fig:sensNMO}). Cross sections have been also studied in more detail by modifying the six most relevant parameters of the model implemented in GENIE. The reduction in sensitivity was found to be negligible. Studies within ORCA and PINGU have tested the impact of $\delta_\mathrm{CP}$ and found an additional reduction of up to 0.5$\sigma$ at the three-year benchmark \cite{ORCA_ICRC, jp_vlvnt}. Note that all figures in this review do not include this effect.

The LLR (and $\Delta \chi^2$ for PINGU) resulting from fits to the pseudo-experiments are used to calculate the separability of the two possible mass orderings. The median (i.e with 50\% statical power) sensitivities to the NMO are shown in Fig~\ref{fig:sensNMO} (top) after 3 years of data taking. The results are obtained by fixing $\delta_{CP}$ to zero, and are shown as a function of $\theta_{23}$. Both collaborations observe that constraining $\theta_{23}$ to either octant while doing a fit artificially increases the sensitivity to the NMO, thus the parameter is left unconstrained in these studies.

Though ORCA and PINGU sensitivities should be compared with caution, as the various inputs are slightly different, both studies find a better sensitivity to the NMO for a true value of $\theta_{23}$ in the second octant in the case of normal mass ordering. For the case of inverted ordering, the sensitivity has a much weaker dependence on the value of $\theta_{23}$. The consistency of the two results is encouraging, as they have been obtained with completely independent analysis chains.

The expected improvement in sensitivities with running time, which does not yet include the effects of $\delta_\mathrm{CP}$ nor the reconstructed inelasticity, are shown in Fig~\ref{fig:sensNMO} (bottom). Once more, the discrimination power of both detectors is comparable.

The identification of the mass ordering devised by both collaborations also produces a measurement of $\theta_{23}$ and the absolute value of the atmospheric mass splitting. Projections of the sensitivity to $\sin^2\theta_{23}$ have a strong dependence on the assumed true values. For $\sin^2\theta_{23}=0.45$ both PINGU and ORCA expect to achieve errors of the order of 0.05 after three years of operation. The precision achievable on the absolute value of the mass splitting is roughly independent of the true value, and the expected error on the measurement for both projects is about $0.05\times 10^{-3}$. Both experiments are expected to produce measurements with better precision to those projected for NOvA and T2K by the year 2020.

The results shown in the present paper are a compilation of the most recent, publicly shown projections of both collaborations, and include most leading systematics effects \cite{ORCA_ICRC, PINGU_ICRC, jp_nufact, jp_vlvnt}. Recently a thorough study of the interplay between the oscillations parameters has been reported in \cite{nmh_stats}, consistent with the recent results from ORCA and PINGU. The authors also introduced uncertainties in the estimated energy and zenith resolutions, as well as additional (conservative) uncorrelated uncertainties. Their results show that after 5 years of data taking, the loss in sensitivity ranges from 24\% to 40\% under pessimistic assumptions (such as fully uncorrelated errors in each analysis bin), thus leaving room for a measurement of the NMO by ORCA and PINGU on a reasonable time scale.
\section{Summary}
\label{sec:summary}
Atmospheric neutrinos are a versatile tool to study neutrino oscillations. This naturally occurring beam covers baselines as large as Earth's diameter, and has an energy range which spans over the regimes of oscillations in vacuum, and with resonant and saturated matter effects. Current VLVNTs, ANTARES and IceCube, can detect neutrinos in the latter regime, and have already produced measurements of the atmospheric oscillation parameters, $\theta_{23}$ and $| \Delta m^2_{32}|$. Constant improvements in the understanding and modeling of the detector and media, as well as more sophisticated data analysis techniques have lead to promising results, which have started to become comparable with those of other, more mature experimental set-ups. 

Proposed VLVNTs, ORCA and PINGU, aim to lower the energy threshold and access the resonant regime, with the goal of measuring the sign of $\Delta m^2_{31}$ and completely determining the neutrino mass ordering. While both projects are on the way of optimizing their detector geometries and/or analysis techniques, current studies are nevertheless mature and indicate that they could provide a significant measurement ($\geq3\sigma$, depending on the true value of $\theta_{23}$) of the neutrino mass ordering after $3-4$ years of operation.
\section*{Acknowledgments}

We are grateful to J.~Brunner, M.~Jongen, J.~Hofest\"adt, W.~Winter, D.~Cowen, S.~B\"oser and T.~DeYoung for useful discussions and clarifications both on the theoretical and experimental aspects covered in this paper.

\end{document}